\documentclass[12pt,psamsfonts,reqno]{amsart}
\theoremstyle{definition}

\theoremstyle{remark}

\newcommand {\la} {\left \langle}
\newcommand {\ra} {\right \rangle}

\newcommand{\vev}[1]{\left\langle #1 \right\rangle}
\newcommand{\ket}[1]{\left |  #1 \right \rangle}
\newcommand{\bra}[1]{\left \langle  #1 \right |}

\newcommand {\CalF} {\mathcal F}

\newcommand {\CalN} {\mathcal N}

\newcommand {\CalX} {\mathcal X}

\newcommand {\BR}   {\mathbb R}
\newcommand {\BZ}   {\mathbb Z}
\newcommand {\BC}   {\mathbb C}

\newcommand{\bY}{\mathbf{Y}}
\newcommand{\msS}{\mathscr{S}}
\newcommand{\si}{\mathsf{i}}
\newcommand{\sn}{\mathsf{n}}

\newcommand{\sA}{\mathsf{A}}

\newcommand{\sS}{\mathsf{S}}
\newcommand{\sV}{\mathsf{V}}
\newcommand{\sT}{\mathsf{T}}
\newcommand{\sX}{\mathsf{X}}
\newcommand{\sY}{\mathsf{Y}}

\newcommand{\frakM}{\mathfrak{M}}

\newcommand{\fq}{\mathfrak{q}}
\DeclareMathOperator{\Tr} {Tr}

\numberwithin{equation}{section}

\usepackage[margin=1in]{geometry}

\usepackage{amsmath}
\usepackage{amssymb}
\usepackage{amsxtra}
\usepackage{amscd}
\usepackage[mathscr]{euscript}
\usepackage{bbm}

\usepackage[colorlinks=true,linkcolor=blue,citecolor=blue,urlcolor=blue]{hyperref}
\usepackage[numbers,sort&compress]{natbib}
\usepackage{hypernat}

\usepackage{iftex}
\ifxetex
        \usepackage{fontspec}
\else
\fi

\usepackage{float}
\usepackage{tikz-cd}
\usetikzlibrary{decorations.pathreplacing}
\usepackage[mathscr]{euscript}
\usepackage{framed}
\definecolor{shadecolor}{gray}{0.85}
\newdimen\tbaselineshift

\newcommand{\rem}[1]{\textcolor{magenta}{{\bf (#1)}}}

\usetikzlibrary{decorations.pathreplacing} 
\usetikzlibrary{decorations.markings}
\tikzset{->-/.style={decoration={
  markings,
  mark=at position .5 with {\arrow{>}}},postaction={decorate}}}

\usepackage[framemethod=TikZ]{mdframed}

\newenvironment{itembox}[1]{\begin{mdframed}[
  roundcorner=5pt,
  skipabove=\topskip,
  frametitleaboveskip=\dimexpr-0.7\baselineskip,
  innertopmargin=\dimexpr-0.25\baselineskip,
  innerbottommargin=\dimexpr0.65\baselineskip,
  frametitle={\tikz{\node[anchor=base,rectangle,fill=white]{\strut #1};}}]
  }
  {%\vspace{.3\baselineskip}
   \end{mdframed}}

\begin{document}

\title{Double quantization of Seiberg--Witten geometry and W-algebras}

\author{Taro Kimura}

\address{%Taro Kimura,
Keio University, Japan}

\begin{abstract} 
 We show that the double quantization of Seiberg--Witten spectral curve for $\Gamma$-quiver gauge theory defines the generating current of W$(\Gamma)$-algebra in the free field realization.
 We also show that the partition function is given as a correlator of the corresponding W$(\Gamma)$-algebra, which is equivalent to the AGT relation under the gauge/quiver (spectral) duality.
\end{abstract}

\maketitle 

\tableofcontents

\parskip=4pt

\section{Introduction and summary}

Recent progress on non-perturbative aspects of supersymmetric gauge theory exhibits interesting connections with various concepts of mathematical physics.
In particular, Seiberg--Witten theory for four-dimensional $\CalN=2$ gauge theory~\cite{Seiberg:1994rs,Seiberg:1994aj} provides a geometrical interpretation for the Coulomb branch moduli space as a base of the algebraic integrable system~\cite{Gorsky:1995zq,Martinec:1995by,Donagi:1995cf,Seiberg:1996nz}.
As a result, the Seiberg--Witten curve can be identified as a classical spectral curve of the affine Toda chain for the pure SYM theory, the $SU(2)$ rational spin chain for $\CalN=2$ SQCD with the fundamental hypermultiplets~\cite{Gorsky:1996hs}.

\vspace{.5em}

The relation between gauge theory and integrable system has various generalizations:

%\vspace{.5em}

%\subsubsection*{Dimensional hierarchy}
\begin{itembox}{Dimensional hierarchy}
Replacing the spacetime on which the gauge theory is defined with five and six dimensional manifolds compactified on $S^1$ and $T^2$, it corresponds to trigonometric and elliptic deformations of rational integrable system~\cite{Nekrasov:1996cz,Nekrasov:2012xe}, as shown in Table~\ref{tab:dimensions}.
The Seiberg--Witten curve for 4d/5d/6d theory coincides with the spectral curve of rational/trigonometric/elliptic integrable system.
\end{itembox}

%\vspace{.5em}

%\subsubsection*{Quiver gauge theory}
\begin{itembox}{Quiver gauge theory}
In the relation to integrable system, %the gauge group rank is translated into the spin chain length. This means that
 change of the gauge symmetry doesn't lead to change of underlying symmetry on the integrable system side.
For this purpose, we need quiver gauge theory labeled by a quiver graph $\Gamma$.
The corresponding symmetry group for integrable system turns out to be Lie group $G_\Gamma$ whose Dynkin diagram is given by $\Gamma$, and the Seiberg--Witten geometry is described by the fundamental characters of $G_\Gamma$ group~\cite{Nekrasov:2012xe}.
\end{itembox}

%\vspace{.5em}

%\subsubsection*{$\Omega$-background and quantization}
\begin{itembox}{$\Omega$-background and quantization}
The four-dimensional spacetime for gauge theory allows an equivariant deformation with $\Omega$-background~\cite{Moore:1997dj,Nekrasov:2002qd}, %which is called the $\Omega$-background,
and its specific limit $(\epsilon_1,\epsilon_2) \to (\hbar, 0)$, called the Nekrasov--Shatashvili (NS) limit, corresponds to quantization of integrable system~\cite{Nekrasov:2009zz,Nekrasov:2009rc}.
Then the Seiberg--Witten curve characterized by the algebraic relation is promoted to difference equation, and this quantum spectral curve is identified as Baxter's TQ-relation for $G_\Gamma$-spin chain, which is also equivalent to the corresponding Bethe ansatz equation~\cite{Nekrasov:2013xda} described by the polynomial equation called the $q$-character~\cite{Frenkel:1998} (Table \ref{tab:quantization}).
\end{itembox}

\vspace{.5em}

\begin{table}[t]
 \begin{tabular}{ccc}\hline\hline
  Gauge theory geometry & Geometric realization & Integrable system
	  \\\hline %& Quantum algebra	  
  $\BR^4$ & IIA string on CY3 & rational \\%& Yangian
  $\BR^4 \times S^1$ & M theory on CY3 $\times \, S^1$ & trigonometric \\%& quantum group
      $\BR^4 \times T^2$ & F theory on CY3 $\times \, T^2$ & elliptic
  \\\hline\hline %& elliptic quantum group
 \end{tabular}
 \\[1em]
 \caption{Gauge theory geometry and IIA/M/F theory realization associated with rational/trigonometric/elliptic integrable systems. See, for example, \cite{Hollowood:2003cv}.}
 \label{tab:dimensions}
\end{table}

%\vspace{1em}

\begin{table}[t]
% \vspace{1em}
 \begin{tabular}{ccc}\hline\hline
  $\Omega$-background & SW geometry & Integrable system\\\hline
  $(\epsilon_1,\epsilon_2) = (0,0)$ & character & classical  \\
  $(\epsilon_1,\epsilon_2) = (\hbar,0)$ & $q$-character & quantum \\
  $(\epsilon_1,\epsilon_2) \neq (0,0)$ (w/ $t$-extention)
  & $qq$-character & doubly quantum (W-algebra) \\\hline\hline
 \end{tabular}
 \\[1em]
 \caption{$\Omega$-background and (double) quantization of Seiberg--Witten geometry described by $G_\Gamma$-character/$q$-character/$qq$-character for $\Gamma$-quiver gauge theory.}
 \label{tab:quantization}
\end{table}

From this point of view, we still have another deformation parameter:
We have used only one of equivariant parameters to quantize the integrable system.
Thus it is natural to ask \textit{what is the underlying algebraic structure for generic $\Omega$-background parametrization.}
It has been recently shown that the polynomial equation still holds for generic $(\epsilon_1, \epsilon_2)$ just by replacing the $q$-character with another algebraic structure, called the $qq$-character~\cite{Nekrasov:2015wsu,Nekrasov:2016qym}.
See also \cite{Bourgine:2015szm,Mironov:2015thk,Kim:2016qqs,Bourgine:2016vsq,Mironov:2016cyq,Mironov:2016yue,Awata:2016riz,Awata:2016mxc,Kimura:2017auj} for its realization in gauge theory and topological string theory.

We proceed with this program to construct W-algebras using quiver gauge theory~\cite{Kimura:2015rgi,Kimura:2016dys,Kimura:2017hez}.
In our construction, in addition to consider generic equivariant parameters $(\epsilon_1, \epsilon_2)$, we incorporate the higher time variables, which deform UV prepotential by the local operators~\cite{Marshakov:2006ii}.
Our claim is as follows:
\begin{shaded}\noindent%\noindent\itshape\ignorespaces
The $qq$-character together with double quantization of Seiberg--Witten spectral curve for the $t$-extended $\Gamma$-quiver gauge theory defines the generating current of W($\Gamma$)-algebra.   
\end{shaded}
\noindent
The doubly quantum spectral curve is an operator relation, in contrast to the classical and quantum curves which are relations between correlation functions.
This double quantization scheme will be discussed in Sec.~\ref{sec:2nd} in more detail.

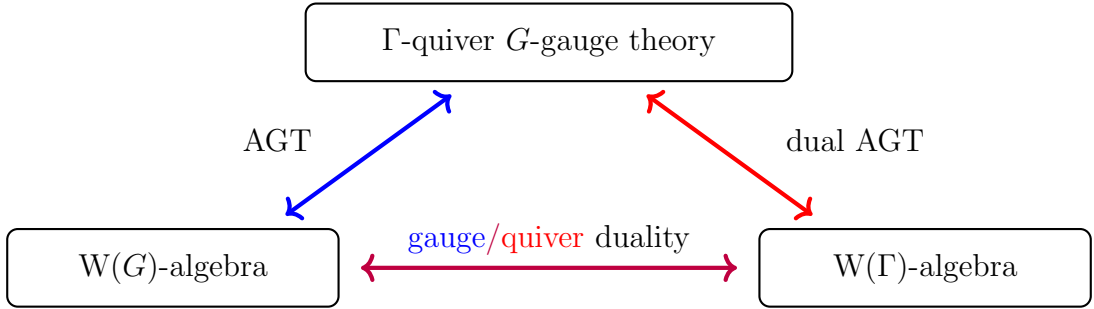
\begin{figure}[t]
 \begin{center}
  \begin{tikzpicture}[thick]

   \node [rectangle, draw, fill=white,%lime!20,
   text width=15em, text centered, rounded corners,
   minimum height=2.5em, %blur shadow={shadow blur steps=5}
   ]
   (4d) at (0,0) {$\Gamma$-quiver $G$-gauge theory};

   \node [rectangle, draw, fill=white,%lime!20,
   text width=10em, text centered, rounded corners,
   minimum height=2.5em, %blur shadow={shadow blur steps=5}
   ]
   (G) at (-5,-3) {W$(G)$-algebra};

   \node [rectangle, draw, fill=white,%lime!20,
   text width=10em, text centered, rounded corners,
   minimum height=2.5em, %blur shadow={shadow blur steps=5}
   ]
   (Gamma) at (5,-3) {W$(\Gamma)$-algebra};   

%   \draw[<->] (4d) -- (G);
%   \draw[<->] (4d) -- (Gamma);
%   \draw[<->] (G) -- (Gamma);

   \draw[<->,ultra thick, blue] (-1.3,-.7) -- (-3.5,-2.3);
   \draw[<->,ultra thick, red] (1.3,-.7) -- (3.5,-2.3);   
   \draw[<->,ultra thick, purple] (-2.5,-3) -- (0,-3)
   node [above] {\textcolor{blue}{gauge}/\textcolor{red}{quiver} \textcolor{black}{duality}} -- (2.5,-3);

   \node at (-3.,-1.3) [left] {AGT};
   \node at (3.,-1.3) [right] {dual AGT};
   
  \end{tikzpicture}
 \end{center}
 \caption{AGT and dual AGT}
 \label{AGT_and_dual}
\end{figure}

Another claim about the partition function is:
\begin{shaded}\noindent
The $t$-extended $\Gamma$-quiver partition function is a state (\emph{$Z$-state}) in the Hilbert space, and the non-$t$-extended partition function is given by a correlation function of W($\Gamma$)-algebra.
\end{shaded}
\noindent
This statement is analogous, but slightly different from the AGT relation, which claims that the partition function of gauge theory enjoying gauge symmetry $G$ is given by a correlation function of W($G$)-algebra~\cite{Alday:2009aq,Wyllard:2009hg}.%
\footnote{%
Our construction naturally gives rise to the $q$-deformation of W($\Gamma$)-algebra from 5d quiver gauge theory, while the ordinary AGT relation claims a connection between 4d gauge theory and W($G$)-algebra without $q$-deformation.
The counterpart of our construction is thus the $q$-AGT relation for 5d gauge theory and the $q$-deformed W($G$)-algebra~\cite{Awata:2009ur}.
A similar idea to realize the Virasoro/W-algebra through quantization of geometry is found in~\cite{Teschner:2010je,Manabe:2015kbj} for non-$q$-deformed case.}
This reflects that our construction of W($\Gamma$) is equivalent to AGT for W($G$) by the gauge/quiver duality $G \leftrightarrow \Gamma$~\cite{Katz:1997eq,Aharony:1997bh,Bao:2011rc}, which is confirmed for $A$-type quiver (Fig.~\ref{AGT_and_dual}).
In the classical limit, this duality becomes the Nahm--Fourier--Mukai duality between the phase space of Hitchin system on a cylinder and the phase space of periodic monopoles~\cite{Nekrasov:2012xe}.
We remark a similar realization of W-algebra correlators associated with the quiver structure~\cite{Aganagic:2013tta,Aganagic:2014oia,Aganagic:2015cta}.

Let us summarize our construction of quiver W-algebras~\cite{Kimura:2015rgi,Kimura:2016dys}.
We start with the classical Seiberg--Witten curve given in the limit $(\epsilon_1,\epsilon_2) \to (0,0)$.
For $\Gamma$-quiver gauge theory, this spectral curve is given by the fundamental characters of $G_\Gamma$ group, if the quiver diagram $\Gamma$ coincides with a Dynkin diagram for finite simple Lie algebras.
This construction is also applicable to generic quiver which doesn't correspond to any finite Lie algebra. %
%\footnote{%
%We assume in this article that the quiver is simply-laced.
%}
Tuning on the equivariant parameter as $(\epsilon_1, \epsilon_2) \to (\hbar, 0)$, we obtain the quantum Seiberg--Witten curve, which is identified as the TQ-relation of the integrable $G_\Gamma$-spin chain.
We then consider the double quantization of Seiberg--Witten curve by taking into account the higher time variables with the fully generic equivariant parameters $(\epsilon_1, \epsilon_2)$.
We obtain an operator relation, which is identified as the quantum Sugawara construction of the generating current of W($\Gamma$)-algebra.

\newpage

This article is a self-contained review based on~\cite{Kimura:2015rgi,Kimura:2016dys}.
Our construction exhibits several new features of W-algebras:

%\subsubsection*{Edge (bifundamental) mass-deformation of W-algebra}
\begin{itembox}{Edge (bifundamental) mass-deformation of W-algebra}
In general, quiver gauge theory consists of the vector multiplet assigned to a node of quiver $i$, and the hypermultiplet in bifundamental representation for an edge $e$ with multiplicative mass parameter $\mu_e$.
(We will deal with (anti)fundamental hypermultiplets separately.)
We will show that the bifundamental mass gives rise to new mass-deformation of W-algebra, which is reduced to Frenkel--Reshetikhin's construction in the limit $\mu_e \to 1$~\cite{Frenkel:1996,Frenkel:1997}.
\end{itembox}

%\vspace{.5em}

\begin{itembox}{Affine quiver W-algebra}
 Our construction of W-algebra is applicable to generic quiver gauge theory.
 Applying this construction to affine quiver theory, e.g., $\CalN = 2^*$ theory ($\widehat{A}_0$-quiver), we obtain new W-algebra associated to affine Lie algebra.
 The Seiberg--Witten curve (and its quantization) is described by the affine character, which is given as an infinite series.
\end{itembox}

%\vspace{.5em}

\begin{itembox}{Quiver elliptic W-algebra}
 We can construct elliptically deformed W-algebras using six-dimensional quiver gauge theory on $\BR^4 \times T^2$, which corresponds to the elliptic integrable system.
 %with the spectral variable $x \in \BC^\times/p^{\BZ}$.
\end{itembox}
This article contains several new features compared with the original papers~\cite{Kimura:2015rgi,Kimura:2016dys}:
We introduce the $\sA$-operator to discuss the iWeyl reflection in Sec.~\ref{sec:quiv_alg}.
We show the explicit OPE relations for $A_2$ quiver in Sec.~\ref{sec:A2}.
We also provide the fundamental $qq$-characters for $D_4$ quiver in Sec.~\ref{sec:Dr}.
Furthermore, generalization to fractional quiver, which gives rise to non-simply-laced algebra, is found in~\cite{Kimura:2017hez}.

\vspace{.5em}

The rest of this article is organized as follows.
We start in Sec.~\ref{sec:1st} to discuss the classical Seiberg--Witten geometry and its 1st quantization.
The resultant quantum spectral curve obtained in the NS limit is a difference operator, which turns out to be Baxter's TQ-relation, and the $Q$-operator plays a role of the wavefunction.
In Sec.~\ref{sec:2nd} we discuss 2nd quantization of the spectral curve.
To obtain an operator relation, we introduce the higher time variable as a Heisenberg oscillator, leading to modification of the UV prepotential.
We then show in Sec.~\ref{sec:qVir} how the generating current of $q$-deformed Virasoro algebra arises in 5d gauge theory.
The construction using the $qq$-character yields the free field realization of the current.
In Sec.~\ref{sec:Zstate} we show that the $t$-extended gauge theory partition function is given by a state, which we call the $Z$-state, in the Hilbert space through the operator/state correspondence.
In particular, the $Z$-state is constructed with the screening charge of the corresponding conformal algebra.
We also show that the (anti)fundamental matter contribution is realized by a vertex operator applied to the state.
In Sec.~\ref{sec:quiv_alg} we show how to construct W($\Gamma$)-algebra associated with generic quiver $\Gamma$ from the mass-deformed $q$-Cartan matrix characterizing the quiver gauge theory.
We discuss the regularity of the $qq$-character which provides the generating current, and show it commutes with the screening charge of W($\Gamma$)-algebra.
We demonstrate our construction with several examples reproducing known results in the limit.
In Sec.~\ref{sec:affine_alg} our construction is applied to affine quiver theory, involving the adjoint matter, corresponding to 4d $\CalN=2^*$ (5d $\CalN=1^*$) theory.
This quiver doesn't correspond to finite Lie algebra, and thus provides a new W-algebra such that the generating current is given by an infinite sum of the fundamental current.
In Sec.~\ref{sec:elliptic_alg} we discuss elliptic generalization of quiver W-algebra using 6d gauge theory defined on $\BR^4 \times T^2$.
The whole construction discussed above is straightforwardly lifted, and thus we obtain elliptically deformed W-algebras associated with generic quiver.

\section{Seiberg--Witten spectral curve and 1st quantization}
\label{sec:1st}

The low-energy effective dynamics of 4d $\CalN=2$ gauge theory is completely determined by the algebraic curve called the Seiberg--Witten curve~\cite{Seiberg:1994rs,Seiberg:1994aj}.
%\begin{align}
% \Sigma =
% \{ (\sX, \sY) \in \BC \times \BC^\times \mid H(\sX, \sY ) = 0 \}
%\end{align}
For $U(n)$ SYM theory it is characterized by the algebraic relation
\begin{align}
 y(x) + \frac{1}{y(x)} & = T_n(x)
 \label{eq:c-curve}
\end{align}
where $T_n(x)$ is a degree $n$ monic polynomial in $x^{-1}$ variable,%
\footnote{%
While we usually use the convention such that $T_n(x)$ is a polynomial in $x$, we apply the opposite convention in this article for latter convenience.
%We will also use the convention so that $T_n(\sX)$ is a polynomial of $\sX^{-1}$ for convenience.
}
and the vev of Coulomb moduli parameter is encoded into its coefficients.%
\footnote{%
In this article we basically use the 5d notation:
\begin{align}
 q_1 = e^{\epsilon_1} \, , \qquad
 q_2 = e^{\epsilon_2} \, , \qquad
 q = q_1 q_2 \, , \qquad
 \{\nu_{i,n}\}_{i \in \Gamma_0,\alpha \in [1,\ldots n_i]}
 = \{e^{a_{i,n}}\}_{i \in \Gamma_0,\alpha \in [1,\ldots n_i]}
 \, .
\end{align}
These parameters correspond to the standard notation of $q$-deformed W-algebras: $(q_1,q_2) = (t^{-1},q)$.
}
This algebraic relation can be obtained from the saddle point analysis of gauge theory partition function~\cite{Nekrasov:2002qd,Nekrasov:2003rj,Nekrasov:2012xe}, and its derivation is quite analogous to the large $N$ analysis of matrix models: $y$-variable is realized as a one-point function of a generating function of gauge invariant observables (the $\sY$-operator)~\cite{Nekrasov:2013xda}
\begin{align}
 y(x) = \Big< \sY(x) \Big>
 \, ,
 \label{eq:Y-op_vev}
\end{align}
which plays a similar role to the resolvent function having a singularity in the complex plane.
Incorporating the one-form differential defined on the curve, called the Seiberg--Witten differential,%
\footnote{%
The higher-dimensional lift is equivalent to imposing periodicity:
The cylindrical coordinate for 5d theory on $\BR^4 \times S^1$ is given by $\BC^\times = \BC/\BZ$.
The geometry for 6d gauge theory compactified on a torus $\BR^4 \times T^2$ is parametrized by a coordinate $(x,y) \in E_{T^2} \times \BC^\times$ where the elliptic curve $E_{T^2}$ is doubly periodic, $E_{T^2} = \BC/(\BZ + \tau \BZ)$ with the torus modulus $\tau$.
}
\begin{align}
 \lambda =
 \begin{cases}
  (4d) \quad
  x \, d \log y
  \quad
  & (x, y) \in \BC \times \BC^\times \\
  (5d) \quad
  \log x \, d \log y
  \quad
  & (x, y) \in \BC^\times \times \BC^\times
 \end{cases}
 \, ,
 \label{eq:SW-1form}
\end{align}
they define the Seiberg--Witten geometry describing the SUSY vacuum structure. 
We remark that the curve given by \eqref{eq:c-curve} coincides with the classical spectral curve of the affine $n$-site Toda chain, and also the inhomogeneous limit of $SU(2)$ rational/trigonometric/elliptic spin chain for 4d/5d/6d gauge theory.
%This is an aspect of the connection between gauge theory and integrable system, which we will discuss in details later.

Although the spectral curve is obtained in the classical limit $\epsilon_1, \epsilon_2 \to 0$ $(q_1, q_2 \to 1)$, a similar structure can be found even after turning on the equivariant parameters.
Taking into account one of the equivariant parameters $(\epsilon_1, \epsilon_2) \to (\hbar, 0)$, which is called the Nekrasov--Shatashvili limit, we obtain a difference equation instead of the algebraic relation for 5d theory~\cite{Nekrasov:2009rc,Nekrasov:2013xda}%
\footnote{%
This is also called the $q$-character due to its representation theoretical origin in quantum affine algebra~\cite{Frenkel:1998}.
}
\begin{align}
 y(x) + \frac{1}{y(q_1^{-1}x)} & = T_n(x)
 \, .
 \label{eq:q-curve}
\end{align}
This is again obtained by the saddle point analysis with respect to $\epsilon_2 \to 0$, and the other finite parameter $\epsilon_1 = \hbar$ plays a role of the Planck constant.
Indeed the difference equation \eqref{eq:q-curve} is interpreted as a quantization of the spectral curve \eqref{eq:c-curve} with respect to a canonical pair $(\log x,\log y)$,
\begin{align}
 [ \log y, \log x] = \hbar
 \, ,
 \label{eq:CCM}
\end{align}
which corresponds to the one-form \eqref{eq:SW-1form} or the associated symplectic two-form
\begin{align}
% d\lambda
 \Omega = d \log x \wedge d \log y
 \, .
\end{align}
This quantization implies the $y$-variable behaves as a shift operator $y = \exp \left( \hbar \partial_{\log x} \right)$, and thus the classical curve \eqref{eq:c-curve} becomes a difference operator
\begin{align}
 \left[
 e^{\hbar \partial_{\log x}}
 + e^{- \hbar \partial_{\log x}}
 - T_n(x)
 \right] Q(x) = 0
 \, ,
 \label{eq:q-curve2}
\end{align}
which is equivalent to \eqref{eq:q-curve} under identification%
\footnote{%
This transformation is analogous to that for the Riccati-type differential equation, widely used to derive quantum spectral curves.
}
\begin{align}
 y(x) = \frac{Q(q_1 x)}{Q(x)}
 \, .
\end{align}
In this way, quantization of spectral curve gives a differential/difference equation.
In particular, in this case, the quantum curve \eqref{eq:q-curve2} is equivalent to Baxter's TQ-relation for $SU(2)$ spin chain,%
\footnote{%
To obtain a precise agreement, we have to take into account the Chern--Simons factor and the (anti)fundamental hypermultiplets.
See~\cite{Nekrasov:2009rc,Nekrasov:2013xda} for more arguments.}
which is a consequence of the connection between gauge theory and integrable system at the quantum level.
%Furthermore, since the saddle point equation \eqref{eq:q-curve} is equivalent to Bethe ansatz equation, such a connection is sometimes called the gauge/Bethe correspondence.

In addition, it has been recently revealed that a similar polynomial relation holds even away from the saddle point analysis, which is called the $qq$-character equation~\cite{Nekrasov:2015wsu,Bourgine:2015szm,Mironov:2016yue,Awata:2016riz,Bourgine:2016vsq}
\begin{align}
 y(x) + \frac{1}{y(q^{-1}x)} & = T_n(x)
 \, .
 \label{eq:qq-curve}
\end{align}
Precisely speaking, the second factor should be $y(q^{-1} x)^{-1} \to \left< \sY(q^{-1}x)^{-1} \right>$ as discussed below.
By the analogy with the matrix model and other related theories, this is quite natural for quantum curve.
For example, in the case of the matrix model, the characteristic polynomial average plays a similar role to the wavefunction $Q(x)$ in \eqref{eq:q-curve2}, and actually satisfies the quantum curve equation at finite $N$ under identification $\hbar = 1/N$.
In general, quantization scheme is a way to go back to the finite coupling regime from the classical limit.
We remark that, in the unrefined limit $q_1 q_2 = q = 1$ ($\epsilon_1 + \epsilon_2 = 0$), it is not distinguishable from the classical relation~\eqref{eq:c-curve}, since the summation plays a role of the Planck constant in this case, $\epsilon_1 + \epsilon_2 = \hbar$.
However, when we consider the higher-weight character, an extra factor has to be taken into account as discussed later.
%However the claim is that we don't need to take the classical limit $(\epsilon_1, \epsilon_2) \neq 0$ to obtain the relation.

\section{Operator formalism and 2nd quantization}
\label{sec:2nd}

As mentioned above, the $y$-variable plays a similar role to the resolvent function, and thus the classical spectral curve~\eqref{eq:c-curve} and its quantizations \eqref{eq:q-curve} and \eqref{eq:qq-curve} yield relations between (one-point) correlation functions in gauge theory.
To emphasize it, let us rewrite the $qq$-character with the $\sY$-operator
\begin{align}
 \Big\langle \sY(x) \Big\rangle
 + \Big\langle \frac{1}{\sY(q^{-1} x)} \Big\rangle
 & =
 T_n(x)
 %\Big\langle T(\sX) \Big\rangle
 \, .
\end{align}
Now the degree of $T_n(x)$ depends on the gauge group rank. % which is not universal.
However, the meaning of average $\la \, \mathcal{O}(x) \, \ra$ actually depends on the theory that we consider (gauge group, matter content, and so on).
From this point of view, it is natural to define $T$-operator whose average with $U(n)$ SYM theory is given by
\begin{align}
 \Big\langle T(x) \Big\rangle = T_n(x)
 \, .
 \label{eq:T-op-red}
\end{align}
Then we obtain a universal operator relation
\begin{align}
 \sY(x) + \frac{1}{\sY(q^{-1} x)} & = T(x)
 \, .
 \label{eq:qq-curve2}
\end{align}
This is the 2nd quantization of Seiberg--Witten spectral curve in our sense, and this equation defines the $T$-operator in terms of the $\sY$-operator.
We remark this operator relation is independent of gauge group rank, in contrast to the polynomial relation.
In this sense, it can be seen as a universal relation for gauge theory.

Precisely speaking, in order to obtain a proper operator relation, we need to take into account the higher time variables, namely the potential term in gauge theory
\begin{align}
 Z_\text{pot}(t) =
% \left<
 \exp \left( \sum_{n=1}^\infty t_n %\mathcal{O}_n
 \bY^{[n]}
 \right)
% \right>
 \, ,
 \label{eq:pot-term}
\end{align}
so that the $t$-extended partition function plays a role of generating function of the gauge theory observable $\bY^{[n]}$, which is a trace of the $n$-th power adjoint scalar $\Tr \Phi^n$ for 4d, or its loop operator along a compactified circle for 5d theory.
Inclusion of the potential term corresponds to deformation of the UV prepotential with the holomorphic operators~\cite{Marshakov:2006ii}
\begin{align}
 \CalF_\text{UV} \ \longrightarrow \
 \CalF_\text{UV} + \sum_{n=1}^\infty t_n \Tr \Phi^{n}
 \, .
\end{align}
Then since the situation is completely parallel with the matrix model, let us explain the reason why we need the potential term \eqref{eq:pot-term} by the analogy.
The resolvent operator of matrix model is a generating function of the gauge invariant observable, which is a single trace operator,
\begin{align}
 \Tr \frac{1}{x - X} 
 & = \sum_{n = 0}^\infty x^{-n-1} \, \Tr X^n
 \, .
\end{align}
This observable is equivalent to derivative with respect to the coupling constant, playing a role of the time variable, in the potential term
\begin{align}
 \frac{\partial}{\partial t_n} \, e^{\Tr V(X)}
 = \Tr X^n \, e^{\Tr V(X)}
\end{align}
where
\begin{align}
 V(x) = \sum_{n=1}^\infty t_n \, x^n
 \, .
\end{align}
In this sense the single trace observable is replaced with the derivative with $t_n$ under the presence of the potential term.

\section{From double quantization to Virasoro/W-algebra}
\label{sec:qVir}

In terms of the gauge theory observable, the $\sY$-operator (in particular, its average) is obtained~\cite{Nekrasov:2012xe,Nekrasov:2013xda}
\begin{align}
 \Big\langle
  \sY(x)
 \Big\rangle
 =
 \exp
 \left(
  - \sum_{n=1}^\infty \frac{x^{-n}}{n} \bY^{[n]}
 \right)
 \, .
\end{align}
Taking into account the higher time variables, the $\sY$-operator shows $t$-dependence, which is written in terms of $(q_1,q_2)$-modified free field operators~\cite{Kimura:2015rgi}
\begin{align}
 \sY(x) & = 
 q_1^{\frac{1}{2}} : \exp
 \left(
  y_0 + \sum_{n \neq 0} y_n \, x^{-n}
 \right):
\end{align}
where the free fields are defined
\begin{align}
 (n>0) \qquad
 y_{-n} =
 \frac{(1-q_1^n)(1-q_2^n)}{1+q^n} \, t_n
 \, , \qquad
 y_{n} =
 - \frac{1}{n} \frac{\partial}{\partial t_n}
 \, , \qquad
 y_0 = - \frac{1}{2} t_0 \log q_2
 \, .
 \label{eq:y-field}
\end{align}
A constant and zero mode factors are added for convenience.
According to the operator relation~\eqref{eq:qq-curve2}, it defines the $T$-operator in the free field realization, and it turns out to coincide with the generating current of $q$-deformed Virasoro algebra~\cite{Shiraishi:1995rp,Frenkel:1996}.
Indeed the generator appearing in the mode expansion
\begin{align}
 T(x) = \sum_{n \in \BZ} T_n \, x^{-n}
 \label{eq:T-op-exp}
\end{align}
satisfies the algebraic relation
\begin{align}
 \left[ T_n, T_m \right]
 =
 - \sum_{k=1}^\infty f_k
 \left(
  T_{n-k} T_{m+k} - T_{m-k} T_{n+k}
 \right)
 - \frac{(1-q_1)(1-q_2)}{1-q} \left( q^n - q^{-n} \right)
 \delta_{n+m,0}
 \label{eq:qVir_alg}
\end{align}
where
\begin{align}
 \sum_{k=0}^\infty f_k \, x^k
 & =
 \exp \left(
 \sum_{n=1}^\infty \frac{1}{n}
 \frac{(1-q_1^n)(1-q_2^n)}{1+q^n} x^n
 \right)
 =: f(x)
 \, .
\end{align}
In practice, it is convenient to see the $q$-analog of OPE from the two-point function of the generating current, which is equivalent to the algebraic relation \eqref{eq:qVir_alg},
\begin{align}
 f\left(\frac{y}{x}\right) T(x) T(y)
 - f\left(\frac{x}{y}\right) T(y) T(x)
 & =
 - \frac{(1-q_1)(1-q_2)}{1-q}
 \left(
 \delta \left(\frac{qy}{x}\right)
 - \delta \left(\frac{qx}{y}\right)
 \right)
 \label{eq:qVir_OPE}
\end{align}
where the delta function is defined
\begin{align}
 \delta(x) = \sum_{n \in \BZ} x^n
 \, .
 \label{eq:delta-func}
\end{align}
This shows that double quantization of Seiberg--Witten spectral curve provides the generating current of $q$-Virasoro algebra in the free field realization.
We remark that such an idea on the realization of generating current through the quantization of Seiberg--Witten curve was already discussed in the context of the (non $q$-deformed) AGT relation~\cite{Alday:2009aq}.

\if0
\rem{To be improved: the meaning of vev., Gaiotto state~\cite{Gaiotto:2009ma}}
Let us comment on the classical limit of the doubly quantum curve.
In order to obtain the classical spectral curve, we need to consider the average of the $T$-operator~\eqref{eq:T-op-red}.
For this purpose, we introduce a shifted operator instead of \eqref{eq:T-op-exp},
\begin{align}
 T^{(k)}(x) & =
 \sum_{m \in \BZ} T_m \, x^{-m-k}
 \, .
\end{align}
This shift corresponds to insertion of Chern--Simons term in gauge theory, and the canonical value of $k$ turns out to be the gauge group rank.
Then, for $U(n)$ SYM theory, its average with a primary vacuum yields a degree $n$ polynomial in $\sX^{-1}$,
\begin{align}
 \Big\langle T^{(n)}(\sX) \Big\rangle
 = \sX^{-n} + \cdots
 = T_n(\sX)
 \, ,
\end{align}
where the $T_0$ vev is normalized $\vev{T_0} = 1$.
\fi

What we should remark is that the conformal algebra obtained here using the gauge theory does not depend on the gauge group rank: we obtain ($q$-deformed) Virasoro algebra for $U(n)$ gauge theory for arbitrary $n \in \BZ_{\ge 1}$.
This situation is actually different from the AGT relation where the underlying conformal algebra is associated to the gauge group.
%We will explain how our construction is related to the AGT relation under the duality.

\section{$Z$-state}
\label{sec:Zstate}

Let us see how the gauge theory partition function is interpreted in our construction.
In particular, the partition function extended with the time variables~\eqref{eq:pot-term} explicitly depends on them.
Since the time variable behaves as an operator, the $t$-extended partition function $Z(t)$ has to be identified as an operator.
This interpretation directly leads to the notion of $Z$-state through the operator/state correspondence of CFT.

We show that the $t$-extended full partition function (including both one-loop and instanton contributions) for $U(n)$ SYM theory is a state given by~\cite{Kimura:2015rgi}
\begin{align}
 \ket{Z} & =
 \sum_{\CalX \in \frakM^\sT} \prod_{x \in \CalX}^\succ
 S(x) \ket{1}
 \label{eq:Z-state1}
\end{align}
where the product is a radial-ordered product, and $\frakM^\sT$ is the torus action fixed point in the instanton moduli space, characterized by a set of partitions~\cite{Nekrasov:2002qd} with
\begin{align}
 \CalX = \{ x_{\alpha,k} \}_{\alpha\in [1\ldots n],k\in [1\ldots \infty]}
 \, , \qquad
 x_{\alpha,k} = \nu_{\alpha} q_1^{k-1} q_2^{\lambda_{\alpha,k}}
 \, .
\end{align}
The vacuum state is denoted by $\ket{1}$, which is a constant with respect to the time variables, $\partial_{t_n} \ket{1} = 0$, namely a primary state, and $S(x)$ is a screening current operator
\begin{align}
 S(x) & = \
 :
 \exp
 \left(
 s_0 \log x + \tilde{s}_0
 + \frac{\kappa}{2} \left( \log^2_{q_2} x - \log_{q_2} x \right)
 + \sum_{n \neq 0} s_n \, x^{-n}
 \right)
 :
 \, .
\end{align}
As well as the $\sY$-operator, they are $(q_1,q_2)$-deformed free field operators
\begin{align}
 (n > 0) \qquad
 s_{-n} = (1-q_1^n) t_n
 \, , \qquad
 s_n = - \frac{1+q^{-n}}{n(1-q_2^{-n})} \frac{\partial}{\partial t_n}
 \, , \qquad
 s_0 = t_0
 \label{eq:s-field}
\end{align}
obeying the commutation relation
\begin{align}
 \Big[ s_n, s_m \Big] & =
 - \frac{1}{n} \frac{1-q_1^n}{1-q_2^{-n}} (1+q^{-n}) \,
 \delta_{n+m,0}
 \, .
\end{align}
The zero mode $t_0$ corresponds to the gauge coupling $t_0 = \log_{q_2} \fq$, but we need some redefinition to obtain precise agreement with the gauge theory result.
The additional zero mode satisfies
\begin{align}
 \Big[ \tilde{s}_0, s_n \Big] = - 2 \beta \delta_{n,0}
\end{align}
where $\beta = - \epsilon_1/\epsilon_2$, and the $\kappa$-term is added to reproduce the Chern--Simons term in gauge theory.
See \cite{Kimura:2015rgi} for details.

The $Z$-state can be also written using the screening charge, which commutes with the generating current $T(x)$.
First of all, the summation over the fixed points $\frakM^\sT$ in the partition function can be replaced with $\BZ^{\CalX_0}$, which is a set of arbitrary integer sequences terminating by zeros.
$\CalX_0$ is the ground configuration $\lambda_{\alpha,k} = 0$ for $k\in\BZ_{>0}$, and $\mathring{x}_{\alpha,k} = \nu_\alpha q_1^{k-1}$.
If the partition $\lambda_{\alpha,k}$ is not a non-increasing sequence, there appears a zero factor in the expression
\begin{align}
 \prod_{x \in \CalX}^\succ S(x) \ket{1} = 0
\end{align}
for $\CalX \in \BZ^{\CalX_0}$ but $\CalX \not \in \frakM^\sT$.
Therefore the $Z$-state is given by
\begin{align}
 \ket{Z} & =
 \prod_{\mathring{x} \in \CalX_0}^\succ
 \sS(\mathring{x}) \ket{1}
 \label{eq:Z-state}
\end{align}
where we define the screening charge%
\footnote{%
This infinite sum can be written as the Jackson integral with respect to $q_2$,
\begin{align}
 \sS(x) & =
 \int dx_{q_2} S(x)
 \, ,
\end{align}
which is more convenient to discuss its convergence property.
In the limit $q_2 \to 1$, the discrete sum is replaced with the usual integral.
}
\begin{align}
 \sS(\mathring{x}) & = \sum_{k \in \BZ} S(q_2^k \mathring{x})
 \, .
 \label{eq:S-charge}
\end{align}
For $q$-deformed Virasoro/W-algebra, the screening charge is a discrete sum of the screening current.
Precisely speaking, this sum is given by the Jackson integral to obtain a proper convergence property.

Let us mention an important property of the screening charge.
The OPE between the $\sY$-operator and the screening current is given by
\begin{align}
 \sY(x) S(x')
 =
 \frac{1 - x'/x}{1 - q_1 x'/x}
 : \sY(x) S(x') :
 \, , \quad
 S(x') \sY(x)
 =
 q_1^{-1}
 \frac{1 - x/x'}{1 - q_1^{-1} x/x'}
 : \sY(x) S(x') :
\end{align}
and thus the commutation relation shows
\begin{align}
 \left[ \sY(x), S(x') \right] & =
 (1 - q_1^{-1}) \, \delta\left( q_1 \frac{x'}{x} \right)
 : \sY(x) S(x') :
\end{align}
where the delta function defined \eqref{eq:delta-func}.
Similarly we obtain
\begin{align}
 \left[ \sY(q^{-1} x)^{-1}, S(x') \right] & =
 - (1 - q_1^{-1}) \, \delta\left( q \frac{x'}{x} \right)
 : \sY(x) S(x') :
\end{align}
where we used the formula
\begin{align}
 q_1^{-1} : \sY(x) \sY(q^{-1} x):
 & = \
 : S(q^{-1} x) S(q_1^{-1} x)^{-1} :
\end{align}
obtained by comparing the $y$ and $s$ fields, \eqref{eq:y-field} and \eqref{eq:s-field}.
We have almost the same OPEs for $\sY$ and $\sY^{-1}$ except for the $q_2$-shift.
This discrepancy is canceled in the screening charge because it is defined as a summation over the $q_2$-shifted screening currents.
Therefore the $T$-operator turns out to commute with the screening charge
\begin{align}
 \Big[ T(x), \sS(x') \Big] & = 0
 \, .
\end{align}
This assures the regularity of the $T$-operator, and the mode expansion \eqref{eq:T-op-exp}.
This commutativity also corresponds to the pole cancellation mechanism in the $qq$-character.

In order to obtain the non-extended plain partition function, we apply a dual vacuum state $\bra{1}$ because it plays a role of the projector into the $t=0$ sector, $\bra{1}t_n = 0$.
Thus the plain partition function is given by a correlator
\begin{align}
 Z(t=0)
 = \vev{1|Z}
 = \bra{1} \prod_{\mathring{x} \in \CalX_0}^\succ
 \sS(\mathring{x}) \ket{1}
 \, .
 \label{eq:Z-correlator}
\end{align}
See also~\cite{Aganagic:2013tta,Aganagic:2014oia,Aganagic:2015cta} for a similar construction using little string theory.%
\footnote{%
In these papers, the authors discuss a correlator which consists of finite screening currents by considering the defect operators.
}
This expression resembles the AGT relation~\cite{Alday:2009aq}, which is the coincidence between the gauge theory partition function and the Liouville/Toda CFT correlator.
% but we remark several points.
In the case of the AGT relation, we need to subtract the $U(1)$ factor to obtain agreement between the instanton partition function and the conformal block.
In this case, on the other hand, we don't need to care about such an extra factor.
We can directly see the agreement.
In addition, as mentioned before, the underlying conformal algebra does not depend on the gauge group.
We will come back to this issue later.

So far we have focused on the vector multiplet contribution.
We can also incorporate the fundamental hypermultiplet in the operator formalism.
Since the fixed point contribution to the gauge theory observable is given by%
\footnote{%
This expression can be interpreted as a $q$-deformation of $n$-th Casimir element
\begin{align}
 C_n & = \frac{1}{n}
 \sum_{k=1}^\infty
 \left(
 \left( \lambda_k - k + \frac{1}{2} \right)^n
 - \left( - k + \frac{1}{2} \right)^n
 \right)
 \, .
\end{align}
}
\begin{align}
 \bY^{[n]} & = (1-q_1^n) \sum_{x \in \CalX} x^n
 \, ,
\end{align}
the potential term~\eqref{eq:pot-term} yields the fundamental matter contribution by the shift of time variables
\begin{align}
 t_n \ \longrightarrow \
 t_n + \frac{1}{n} \frac{1}{(1-q_1^n)(1-q_2^n)} \, \mu^{-n}
\end{align}
where $\mu \in \BC^\times$ is the multiplicative mass parameter.
We introduce an operator which induces the time shift
\begin{align}
 \sV(x) = \
 :\exp \left(
 \sum_{n \neq 0} v_n \, x^{-n}
 \right)
 :
\end{align}
where
\begin{align}
 (n>0) \qquad
 v_{-n} = - \frac{1}{1+q^n} t_n
 \, , \qquad
 v_n = \frac{1}{n} \frac{1}{(1-q_1^n)(1-q_2^n)}
 \frac{\partial}{\partial t_n}
 \, .
\end{align}
We compute the OPE of $\sV$ and $S$ operators
\begin{align}
 \sV(x) S(x')
 =
 \left(
  \frac{x'}{x};q_2
 \right)_\infty^{-1}
 :\sV(x) S(x'):
 \, ,
 \qquad
 S(x') \sV(x)
 & =
 \left(
  \frac{q_2 x}{x'};q_2
 \right)_\infty
 :\sV(x) S(x'):
 \, .
\end{align}
They correspond to the fundamental and antifundamental hypermultiplet contributions, while the OPE of $\sV$ and $\sV$ does not yield dynamical contribution.
Thus the extended partition function in the presence of (anti)fundamental matters is obtained by inserting the $\sV$-operators
\begin{align}
 \ket{Z} & =
 \left( \prod_{x \in \CalX_\text{f}} \sV(x) \right)
 \left(
 \prod_{\mathring{x} \in \CalX_0}^\succ
 \sS(\mathring{x}) 
 \right)
 \left( \prod_{x \in \tilde\CalX_\text{f}} \sV(x) \right) 
 \ket{1}
\end{align}
where $\CalX_\text{f} = \{\mu_{f}\}_{f\in[1\ldots n^\text{f}]}$ and $\tilde\CalX_\text{f} = \{\tilde\mu_{f}\}_{f\in[1\ldots \tilde\sn^\text{f}]}$ are sets of fundamental and antifundamental mass parameters.
This $\sV$-operator puts a singularity on the spectral curve at $x = \mu_{f}$.
Then the plain partition function is given by a correlator with additional vertex operators
\begin{align}
 Z(t=0) & =
 \bra{1}
 \left( \prod_{x \in \CalX_\text{f}} \sV(x) \right)
 \left(
 \prod_{\mathring{x} \in \CalX_0}^\succ
 \sS(\mathring{x}) 
 \right)
 \left( \prod_{x \in \tilde\CalX_\text{f}} \sV(x) \right)
 \ket{1}
 \, .
\end{align}
This correlator is given by an infinite discrete sum, which can be interpreted as a $q$-analog of Dotsenko--Fateev's integral formula of the conformal block~\cite{Dotsenko:1984nm,Dotsenko:1984ad}.

\section{Quiver W-algebra}
\label{sec:quiv_alg}

The construction shown above is applicable to generic quiver gauge theory, and we correspondingly obtain the W-algebra associated to the quiver structure.
Let us first fix the notations.
Let $\Gamma$ be a quiver with the set of nodes $\Gamma_0$ and the set of edges $\Gamma_1$.
A quiver $\Gamma$ defines a $|\Gamma_0|\times|\Gamma_0|$ matrix
\begin{align}
 c_{ij}^{[n]} & =
 (1 + q^{-n}) \delta_{ij}
 - \sum_{e: i \to j} \mu_e^{-n}
 - \sum_{e: j \to i} \mu_e^n q^{-n}
\end{align}
where $\mu_e \in \BC^\times$ is the multiplicative bifundamental mass parameter associated to each edge $e \in \Gamma_1$.
This is $(q,\mu_e)$-deformation of Cartan matrix, reduced to the ordinary Cartan matrix in the limit $n \to 0$.
In the analogy with matrix model, the quiver theory with gauge group $\times_{i \in \Gamma_0} U(n_i)$ corresponds to the ADE multi-matrix model~\cite{Kharchev:1992iv,Kostov:1992ie}, so that we have $|\Gamma_0|$ sets of {\it eigenvalues}
\begin{align}
 \CalX_i = \{ x_{i,\alpha,k} \}_{ \alpha \in [1\dots \sn_i], k \in
 [1 \dots \infty] }, \qquad
 \CalX = \bigsqcup_{i \in \Gamma_0} \CalX_i
 \, .
\end{align}
Let $\si$: $\CalX \to \Gamma_0$ be the node label so that $\si(x) = i$ for $x \in \CalX_i$.

Then the construction is quite parallel with the previous situation: The $t$-extended partition function is given by a state
\begin{align}
 \ket{Z} & =
 \prod_{\mathring{x} \in \CalX_0}^\succ
 \sS_{\si(\mathring{x})}(\mathring{x})
 \ket{1}
 \, .
 \label{eq:Z-state-quiv}
\end{align}
The screening charge is defined as a summation over the screening current defined
\begin{align}
 S_i(x) & = \
 :
 \exp
 \left(
 s_{i,0} \log x + \tilde{s}_{i,0}
 + \frac{\kappa_i}{2} \left( \log^2_{q_2} x - \log_{q_2} x \right)
 + \sum_{n \neq 0} s_{i,n} \, x^{-n}
 \right)
 :
 \, .
\end{align}
where the $(q_1,q_2)$ modified free fields are written using the deformed Cartan matrix
\begin{align}
 (n > 0) \qquad
 s_{i,-n} = (1-q_1^n) t_{i,n}
 \, , \qquad
 s_{i,n} = - \frac{1}{n(1-q_2^{-n})} \, c_{ji}^{[n]}
 \frac{\partial}{\partial t_{j,n}}
 \, , \qquad
 s_{i,0} = t_{i,0}
\end{align}
which obey the commutation relation
\begin{align}
 \Big[ s_{i,n}, s_{j,m} \Big] & =
 - \frac{1}{n} \frac{1-q_1^n}{1-q_2^{-n}} \, c_{ji}^{[n]} \,
 \delta_{n+m,0}
 \, .
\end{align}
Again the zero mode is related to the gauge coupling $t_{i,0} = \log_{q_2} \fq_i$, under some redefinition.
The additional zero mode similarly satisfies
\begin{align}
 \Big[ \tilde{s}_{i,0}, s_{j,n} \Big] & =
 - \beta \, \delta_{n,0} \, c_{ji}^{[0]}
 \, .
\end{align}
We can show that the $Z$-state \eqref{eq:Z-state-quiv} coincides wit the $t$-extended gauge theory partition function~\cite{Kimura:2015rgi}.

Let us then discuss how the generating current is constructed using the $\sY$-operator.
For quiver gauge theory, the $\sY$-operator is defined for each node $i \in \Gamma_0$, corresponding to the fundamental weight,
\begin{align}
 \sY_i(x) & = q_1^{\tilde{\rho}_i}
 : \exp \left(
  y_{i,0} + \sum_{n \neq 0} y_{i,n} \, x^{-n}
 \right):
\end{align}
where $\tilde{\rho}_i = \sum_{j \in \Gamma_0} \tilde{c}_{ji}^{[0]}$ is the Weyl vector for non-affine quiver, and $\tilde{\rho}_i = 0$ for affine quiver.
The modified free fields are given by%
\footnote{%
For the quiver whose Cartan matrix is not invertible $\det c^{[0]} = 0$, we cannot incorporate the coupling constant into the zero mode $t_{i,0} = \log_{q_2} \fq_i$.
One needs to put the $\fq_i$ factor for every iWeyl reflection by hand.
}
\begin{align}
 (n > 0) \quad
 y_{i,-n} = (1-q_1^n)(1-q_2^n) \, \tilde{c}_{ji}^{[-n]} t_{j,n}
 \, , \quad
 y_{i,n} = - \frac{1}{n} \frac{\partial}{\partial t_{i,n}}
 \, , \quad
 y_{i,0} = - \tilde{c}_{ji}^{[0]} t_{j,0} \log q_2
% \, .
 \label{eq:y-field-quiv}
\end{align}
with the commutation relation
\begin{align}
 \Big[ y_{i,n} , y_{j,m} \Big]
 & =
 - \frac{1}{n} (1-q_1^n)(1-q_2^n) \tilde{c}_{ij}^{[-n]} \delta_{n+m,0}
 \nonumber \\[.5em]
 & =
 - \frac{1}{n} (1-q_1^{-n})(1-q_2^{-n}) \tilde{c}_{ji}^{[n]} \delta_{n+m,0} 
 \, .
\end{align}
In this case, we can define the $T$-operator which starts with each operator $\sY_i(x)$ for $i \in \Gamma_0$.
It is generated by the local reflection, which is the mass deformed iWeyl reflection~\cite{Nekrasov:2015wsu}%
\footnote{%
In the classical limit $q_1,q_2,\mu_e \to 1$, this reflection is reduced to the ordinary Weyl reflection, and thus the classical Seiberg--Witten geometry is expressed using the character of group $G_\Gamma$, associated to Dynkin-quiver diagram $\Gamma$~\cite{Nekrasov:2012xe}.
}
\begin{align}
 \sY_i(x)
 \ \longrightarrow \
 \sY_i(x) \times
 \left(
 \frac{1}{\displaystyle \sY_i(x) \sY_i(q^{-1} x)}
 \prod_{e:i \to j} \sY_j(\mu_e^{-1} x) 
 \prod_{e:j \to i} \sY_j(\mu_e q^{-1} x)
 \right)
\end{align}
where we assume that there is no self-connecting edge in the quiver.
See Sec.~\ref{sec:affine_alg} for $\widehat{A}_0$ quiver.
This reflection formula is obtained by comparing configurations at the fixed points which differ by one instanton factor.
This means that we shall see the behavior under the partition shift $\lambda_{i,\alpha,k} \to \lambda_{i,\alpha,k}+1$, which is equivalent to $x \to q_2 x$ for $x \in \CalX_i$.
Although the $\sY$-operator itself has a singularity, it is cancelled by the reflection.
Then we obtain
\begin{align}
 T_i(x) & =
 \sY_i(x)
 \, +
 : \sY_i(q^{-1} x)^{-1}
 \prod_{e:i \to j} \sY_j(\mu_e^{-1} x) 
 \prod_{e:j \to i} \sY_j(\mu_e q^{-1} x) :
 + \cdots
 \, .
 \label{eq:iWeyl_ref}
\end{align}
If the reflection gives rise to another singularity, we apply another reflection to cancel it.
For the finite-type quiver, this procedure closes within a finite time.

Since the $Z$-state \eqref{eq:Z-state-quiv} is constructed by the screening charges, we consider the shift of screening current,
\begin{align}
 :\frac{S_i(x)}{S_i(q_2 x)}: & = q_1^{-1} \sA_i(x)
\end{align}
where the $\sA$-operator is defined
\begin{align}
 \sA_i(x) & = \, q_1
 :\exp
 \left(
 a_{i,0}
% - \kappa_i \log x
 + \sum_{n \neq 0} a_{i,n} \, x^{-n}
 \right):
\end{align}
with the free field operators
\begin{align}
 a_{i,n} = (1-q_2^{-n}) s_{i,n}
 \, , \qquad
 a_{i,0} = - t_{i,0} \log q_2
 \, .
\end{align}
These $\sA$ and $\sY$ operators correspond to ``root'' and ``fundamental weight'' respectively~\cite{Frenkel:1997}.
Thus the oscillators are related as follows:
\begin{align}
 a_{i,n} & = y_{j,n} \, c_{ji}^{[n]}
 \, ,
\end{align}
or explicitly,
\begin{align}
 \sA_i(x) & = \, %q_1^{-1}
 :\frac{\sY_i(x) \sY_i(qx)}
      {\displaystyle
       \prod_{e:i \to j} \sY_j(\mu_e^{-1} q x)
       \prod_{e:j \to i} \sY_j(\mu_e x)
      } :
 \, .
\end{align}
This leads to the pole cancellation in the iWeyl reflection
\begin{align}
 \underset{x'\to q^{-1}x}{\operatorname{Res}}
 \Big[ \sY_i(x) S_i(q_2 x') \Big]
 + \underset{x' \to q^{-1} x}{\operatorname{Res}}
 \Big[
 :\sY_i(x) \sA_i(q^{-1} x)^{-1}:
% \sY_i(q^{-1}x)^{-1}
% \prod_{e:j \to i} \sY_j(\mu_e q^{-1} x)
% \prod_{e:i \to j} \sY_j(\mu_e^{-1} x)
 S_i(x')
 \Big]
 = 0
 \, .
\end{align}
Similarly we can show that the $T$-operator commutes with the screening charge defined as a summation over the screening operator \eqref{eq:S-charge},
\begin{align}
 \Big[ T_i(x), \sS_j(x') \Big] & = 0
 \, .
\end{align}
This shows that the $T$-operator yields the generating current of W($\Gamma$)-algebra associated to $\Gamma$-quiver in the free field realization.

We remark that when there appears a product of the $\sY$-operators from the same node $i \in \Gamma_0$ in the reflection \eqref{eq:iWeyl_ref}, one needs an extra structure
\begin{align}
 :\sY_i(x) \sY_i(x'):
 & + \,
 \msS\left(\frac{x'}{x}\right)
 : \frac{\sY_i(x) \sY_i(x')}{\sA_i(q^{-1} x)} :
 + \, \msS\left(\frac{x}{x'}\right)
 : \frac{\sY_i(x) \sY_i(x')}{\sA_i(q^{-1} x')} :
 + : \frac{\sY_i(x) \sY_i(x')}{\sA_i(q^{-1} x) \sA_i(q^{-1} x')} :
 \label{eq:collision1}
\end{align}
%\begin{align}
% :\sY_i(x) \sY_i(z):
% & + \, \msS(z/x)
% : \sY_i(z) \sY_i(q^{-1} x)^{-1}
% \prod_{e:j \to i} \sY_j(\mu_e q^{-1} x)
% \prod_{e:i \to j} \sY_j(\mu_e^{-1} x) :
% \nonumber \\
% & 
% + \, \msS(x/z)
% : \sY_i(x) \sY_i(q^{-1} z)^{-1}
% \prod_{e:j \to i} \sY_j(\mu_e q^{-1} z)
% \prod_{e:i \to j} \sY_j(\mu_e^{-1} z) :
% \nonumber \\
% &
% +
% :\sY_i(q^{-1}x)^{-1} \sY_i(q^{-1}z)^{-1}
% \prod_{e:j \to i} \sY_j(\mu_e q^{-1} x) \sY_j(\mu_e q^{-1} z)
% \prod_{e:i \to j} \sY_j(\mu_e^{-1} x) \sY_j(\mu_e^{-1} z) :
% \label{eq:collision1}
%\end{align}
where the factor corresponding to the OPE between $\sY$ and $\sA$ operators defined
\begin{align}
 \msS(x) & = \frac{(1-q_1 x)(1-q_2 x)}{(1-q x)(1-x)}
 \, .
 \label{eq:S-factor}
\end{align}
Furthermore, in the collision limit $x' \to x$, it involves a derivative term
\begin{align}
 :\sY_i(x)^2: \,
 & + 
 :\left(
 \mathfrak{c}(q_1,q_2) - \frac{(1-q_1)(1-q_2)}{1-q}
 \partial_{\log x} \log \sA_i(q^{-1} x)
 \right) \frac{\sY_i(x)^2}{\sA_i(q^{-1} x)}:
 + \,
 :\frac{\sY_i(x)^2}{\sA_i(q^{-1} x)^2}:
 \label{eq:collision2} 
\end{align}
%\begin{align}
% :\sY_i(x)^2: \,
% & +
% \left(
% \mathfrak{c}(q_1,q_2) - \frac{(1-q_1)(1-q_2)}{1-q}
% \partial_{\log x} \log \sA_i(q^{-1} x)
%% \left(
%% \frac{\sY_i(x)\sY_i(q^{-1}x)}
%%      {\displaystyle
%%       \prod_{e:j \to i} \sY_j(\mu_e q^{-1} x)
%%       \prod_{e:i \to j} \sY_j(\mu_e^{-1} x)}
%% \right)
% \right)
% \nonumber \\
% & \quad \times
% : \frac{\sY_i(x)}{\sY_i(q^{-1}x)}
% \prod_{e:j \to i} \sY_j(\mu_e q^{-1} x)
% \prod_{e:i \to j} \sY_j(\mu_e^{-1} x) :
% \nonumber \\
% & +
% :
% \left(
% \sY_i(q^{-1}x)^{-1}
% \prod_{e:j \to i} \sY_j(\mu_e q^{-1} x)
% \prod_{e:i \to j} \sY_j(\mu_e^{-1} x)
% \right)^2
% :
% \label{eq:collision2} 
%\end{align}
where the constant defined
\begin{align}
 \mathfrak{c}(q_1,q_2)
 & =
 \lim_{x \to 1} \left( \msS(x) + \msS(x^{-1}) \right)
 = \frac{1-6q+q^2+(q_1+q_2)(1+q)}{(1-q)^2}
 \, .
\end{align}
In the NS limit $q_2 \to 1$, the derivative term vanishes, and the factor becomes $\mathfrak{c}(q_1,1) = 2$ with $\msS(x) \to 1$.
Thus the reflection structure becomes simple.
In general, one can consider the product of more than two $\sY$-operators.
In that case, we obtain correspondingly higher derivative terms.

We notice that the free fields \eqref{eq:y-field-quiv} explicitly depend on the bifundamental mass parameter, which may affect the algebraic relation of W-algebra.
Since our construction coincides with the definition of W-algebra given by~\cite{Frenkel:1997} in the mass less limit $\mu_e \to 1$ for $e \in \Gamma_1$, the algebra obtained from quiver gauge theory yields a new mass deformation of W-algebra.
Furthermore, the bifundamental mass can be absorbed by redefinition of the Coulomb moduli parameter when there is no loop edge~\cite{Nekrasov:2012xe,Nekrasov:2013xda}.
%This implies that the bifundamental mass deformation could give an isomorphic map, which is not obvious at the algebraic level.

\subsection{$A_1$ quiver}
\label{sec:A1}

We show more explicit examples.
The simplest example is the $A_1$ quiver.
The $T$-operator is given by
\begin{align}
 T_1(x) & =
 \sY_1(x) + \frac{1}{\sY_1(q^{-1}x)}
 \, ,
\end{align}
which is a double quantum deformation of the fundamental representation character for $G_\Gamma = SU(2)$
\begin{align}
 \chi^{SU(2)}_{\textbf{2}}(y) = y + y^{-1}
 \, .
\end{align}
In addition, we compute the degree $n$ current, corresponding to $(n/2)$-spin representation of $SU(2)$, with the weight $(w_1, \ldots, w_n)$
\begin{align}
 T_1^{[\mathbf{w}]}(x) & = \,
 :\sY_1(w_1 x) \sY_1(w_2 x) \cdots \sY_1(w_n x): + \cdots
 \nonumber \\
 & =
 \sum_{I \cup J = \{1 \ldots n\}}
 \prod_{i \in I, j \in J}
 \msS \left( \frac{w_i}{w_j} \right)
 :
 \prod_{i \in I} \sY_1(w_i x)
 \prod_{j \in J} \sY_1(w_j q^{-1} x)^{-1}
 :
 \, .
\end{align}
The degree two current is especially used to define the algebraic relation for the $q$-Virasoro algebra~\eqref{eq:qVir_OPE}.

\subsection{$A_2$ quiver}
\label{sec:A2}

The next example is the $A_2$ quiver, consisting of two nodes and an edge connecting them.
We have two $T$-operators
\begin{align}
 T_1(x) & =
 \sY_1 (x) \, +
 :\frac{\sY_2(\mu^{-1}x)}{\sY_1(q^{-1}x)}:
 + \, \frac{1}{\sY_2(\mu^{-1} q^{-1} x)}
 \, , \\
 T_2(x) & =
 \sY_2 (x) \, +
 :\frac{\sY_1(\mu q^{-1} x)}{\sY_2(q^{-1}x)}:
 + \, \frac{1}{\sY_1(\mu q^{-2} x)}
 \, ,
\end{align}
where $\mu = \mu_{1 \to 2} = \mu_{2 \to 1}^{-1} q$.
They correspond to the fundamental and antifundamental representation characters of $G_\Gamma = SU(3)$
\begin{align}
 \chi^{SU(3)}_{\textbf{3}}(y_1,y_2)
 = y_1 + y_1^{-1} y_2 + y_2^{-1}
 \, , \qquad
 \chi^{SU(3)}_{\bar{\textbf{3}}}(y_1,y_2) = y_2 + y_1 y_2^{-1} + y_1^{-1}
 \, .
\end{align}
The OPEs for these $T$-operators are given by
\begin{align}
 &
 f_{11} \left(\frac{y}{x}\right) T_1(x) T_1(y)
 - f_{11} \left(\frac{x}{y}\right) T_1(y) T_1(x)
 \nonumber \\
 & \hspace{8em}
 =
 - \frac{(1-q_1)(1-q_2)}{1-q}
 \left(
 \delta\left(q\frac{y}{x}\right) T_2(\mu^{-1} x)
 - \delta\left(q^{-1}\frac{y}{x}\right) T_2(\mu^{-1} q x)
 \right)
 \\[.5em]
 &
 f_{12}\left(\frac{y}{x}\right) T_1(x) T_2(y)
 - f_{21}\left(\frac{x}{y}\right) T_2(y) T_1(x)
  \nonumber \\
 & \hspace{8em}
 =
 - \frac{(1-q_1)(1-q_2)}{1-q}
 \left(
 \delta\left(\mu q \frac{y}{x}\right)
 - \delta\left(\mu q^{-2}\frac{y}{x}\right)
 \right)
 \\[.5em]
 &\
 f_{22} \left(\frac{y}{x}\right) T_2(x) T_2(y)
 - f_{22} \left(\frac{x}{y}\right) T_2(y) T_2(x)
 \nonumber \\
 & \hspace{8em} 
 =
 - \frac{(1-q_1)(1-q_2)}{1-q}
 \left(
 \delta\left(q\frac{y}{x}\right) T_1(\mu q^{-1} x)
 - \delta\left(q^{-1}\frac{y}{x}\right) T_1(\mu x)
 \right)
\end{align}
where
%\begin{align}
% \sY_i(x) \sY_j(y)
% & =
% f_{ij}\left(\frac{y}{x}\right)^{-1}
% : \sY_i(x) \sY_j(y) :
%\end{align}
%with
\begin{align}
 f_{ij} (x) & =
 \exp
 \left(
 \sum_{n=1}^\infty \frac{(1-q_1^n)(1-q_2^n)}{n} \,
 \tilde{c}_{ij}^{[-n]} x^n
 \right)
 \, .
\end{align}
Now the symmetry under exchange $\sY_1 \leftrightarrow \sY_2$ is obvious.
We can read off the algebraic relation for the $(q,\mu)$-deformed W($A_2$) algebra from these OPEs, which explicitly depends on the mass parameter $\mu$.
We remark that the diagonal element $f_{ii}(x)$ does not depend on the mass parameter $(\mu_e)_{e \in \Gamma_1}$ in general.
It appears only in the off-diagonal element $f_{ij}(x)$ for $i \neq j$.

\subsection{$A_r$ quiver}
\label{sec:Ar}

Let us demonstrate our construction with the linear quiver theory $\Gamma = A_r$.
The $T$-operators are computed using the iWeyl reflection
\begin{align}
 T_i(\mu_{i}^{-1} x) & =
 \sum_{1 \le j_1 \le \cdots \le j_i \le r+1}
 : \prod_{k=1}^i \Lambda_{j_k}(q^{-i+k} x) :
 \label{eq:T-op_Ar}
\end{align}
where
\begin{align}
 \Lambda_i(x) & =
 \sY_{i}(\mu_{i}^{-1} x)
 \sY_{i-1}(\mu_{i-1}^{-1} q^{-1} x)^{-1}
\end{align}
with $\sY_0(x) = \sY_{r+1}(x) = 1$, and the mass product defined
\begin{align}
 \mu_{i} & :=
 \mu_{1 \to 2} \mu_{2 \to 3} \cdots \mu_{i-1 \to i}
 \, .
\end{align}
We remark that these $T$-operators are given by $(q,\mu)$-deformation of fundamental representation characters for $G_\Gamma = SU(r+1)$.
This construction coincides with the quantum Miura transformation for the $q$-deformed W($A_r$)-algebra in the massless limit $\mu_e \to 1$~\cite{Frenkel:1996,Awata:1995zk}

\subsection{$D_r$ quiver}
\label{sec:Dr}

\begin{figure}[t]
 \begin{center}
  \begin{tikzpicture}[thick]
   
   \draw (180:2) circle [radius = .17] node (node1) {};
   \draw (0,0) circle [radius = .17] node (node2) {};
   \draw (40:2.3) circle [radius = .2] node (node3) {};
   \draw (-40:2.3) circle [radius = .2] node (node4) {};   

   \draw [->-] (node2) -- (node1);
   \draw [->-] (node2) -- (node3);
   \draw [->-] (node2) -- (node4);

   \node at (-2.5,0) {$1$};
   \node at (-.2,-.6) {$2$};
   \node at (2.3,1.8) {$3$};
   \node at (2.3,-1.8) {$4$};

%   \node at (-1,.5) {$\mu_{1 \to 2}$};
%   \node at (.6,1.2) {$\mu_{3 \to 2}$};
%   \node at (.6,-1.2) {$\mu_{4 \to 2}$};

   \node at (-1,.5) {$\mu_{1}$};
   \node at (.6,1.2) {$\mu_{3}$};
   \node at (.6,-1.2) {$\mu_{4}$};   
   
  \end{tikzpicture}
 \end{center}
 \caption{$D_4$ quiver decorated with bifundamental mass $\mu_i = \mu_{2 \to i}$.}
 \label{fig:D4quiver}
\end{figure}
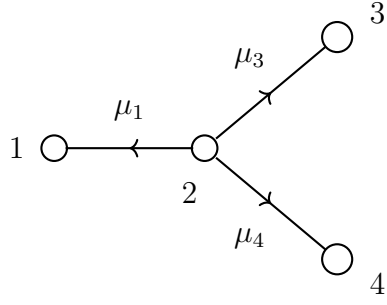

We consider $D$-type quiver gauge theory.
The simplest case is $\Gamma = D_4$, shown in Fig.~\ref{fig:D4quiver}. This quiver shows a symmetry under exchange $1 \leftrightarrow 3 \leftrightarrow 4$, which is known as the $SO(8)$ triality.
Putting the bifundamental mass parameters $\mu_i := \mu_{2 \to i}$ for $i = 1, 3, 4$, we compute the $T$-operator
\begin{align}
 T_1(x) & = \sY_1(x) 
 \, + : \frac{\sY_2(\mu_1 q^{-1} x)}{\sY_1(q^{-1} x)} :
 + : \frac{\sY_3(\mu_1 \mu_3^{-1} q^{-1} x)
           \sY_4(\mu_1 \mu_4^{-1} q^{-1} x)}{\sY_2(\mu_1 q^{-2} x)} :
 + : \frac{\sY_4(\mu_1 \mu_4^{-1} q^{-1} x)}
          {\sY_3(\mu_1 \mu_3^{-1} q^{-2} x)} :
 \nonumber \\[.5em]
 & \quad 
 +
 : \frac{\sY_3(\mu_1 \mu_3^{-1} q^{-1} x)}
        {\sY_4(\mu_1 \mu_4^{-1} q^{-2} x)} :
 +
 : \frac{\sY_2(\mu_1 q^{-2} x)}
        {\sY_3(\mu_1 \mu_3^{-1} q^{-2} x)
         \sY_4(\mu_1 \mu_4^{-1} q^{-2} x)}:
 +
 : \frac{\sY_1(q^{-2} x)}{\sY_2(\mu_1 q^{-3} x)} :
 + \, \frac{1}{\sY_1(q^{-3} x)}
 \, .
\end{align}
The operators $T_3(x)$ and $T_4(x)$ are similarly obtained by permutation.
These three $T$-operators correspond to three $\mathbf{8}$-representations of $SO(8)$.
The remaining $T_2(x)$ operator, corresponding to $\mathbf{28}$-representation, involves collision and derivative terms, \eqref{eq:collision1} and \eqref{eq:collision2},
\begin{align}
 T_2(x) & =
 T_2^+(x) + T_2^-(x)
 +
 \msS(q)
 \left(
 :\frac{\sY_1(\mu_1^{-1} x)}{\sY_1(\mu_1^{-1} q^{-1} x)}:
 + :\frac{\sY_3(\mu_3^{-1} x)}{\sY_3(\mu_3^{-1} q^{-1} x)}:
 + :\frac{\sY_4(\mu_4^{-1} x)}{\sY_4(\mu_4^{-1} q^{-1} x)}: 
 \right)
 \nonumber \\[.5em]
 &
 +
 \left(
 \mathfrak{c}(q_1,q_2) - \frac{(1-q_1)(1-q_2)}{1-q}
 \partial_{\log x} \log
 \left(
 \frac{\sY_2(q^{-1} x) \sY_2(q^{-2} x)}
       {\sY_1(\mu_1^{-1} q^{-1} x)
        \sY_3(\mu_3^{-1} q^{-1} x)
        \sY_4(\mu_4^{-1} q^{-1} x)}
 \right)
 \right)
 :\frac{\sY_2(q^{-1} x)}{\sY_2(q^{-2} x)}:
\end{align}
where
\begin{align}
 T_2^{+}(x) & =
 \sY_2(x) \,
 +
 :\frac{\sY_1(\mu_1^{-1}x) \sY_3(\mu_3^{-1}x) \sY_4(\mu_4^{-1}x)}
       {\sY_2(q^{-1}x)}:
 +
 :\frac{\sY_3(\mu_3^{-1} x) \sY_4(\mu_4^{-1}x)}
       {\sY_1(\mu_1^{-1} q^{-1} x)}:
 +
 :\frac{\sY_1(\mu_1^{-1} x) \sY_4(\mu_4^{-1}x)}
       {\sY_3(\mu_3^{-1} q^{-1} x)}:
 \nonumber \\[.5em]
 &
 +
 :\frac{\sY_1(\mu_1^{-1} x) \sY_3(\mu_3^{-1}x)}
       {\sY_4(\mu_4^{-1} q^{-1} x)}: 
 +
 :\frac{\sY_1(\mu_1^{-1} x) \sY_2(q^{-1} x)}
       {\sY_3(\mu_3^{-1} q^{-1} x) \sY_4(\mu_4^{-1} q^{-1} x)}: 
 +
 :\frac{\sY_3(\mu_3^{-1} x) \sY_2(q^{-1} x)}
       {\sY_1(\mu_1^{-1} q^{-1} x) \sY_4(\mu_4^{-1} q^{-1} x)}: 
 \nonumber \\[.5em]
 &
 +
 :\frac{\sY_4(\mu_4^{-1} x) \sY_2(q^{-1} x)}
       {\sY_1(\mu_1^{-1} q^{-1} x) \sY_3(\mu_3^{-1} q^{-1} x)}: 
 +
 :\frac{\sY_2(q^{-1}x)^2}
       {\sY_1(\mu_1^{-1} q^{-1} x)
        \sY_3(\mu_3^{-1} q^{-1} x)
        \sY_4(\mu_4^{-1} q^{-1} x)}:
 \nonumber \\[.5em]
 &
 +
 :\frac{\sY_1(\mu_1^{-1} x)\sY_1(\mu_1^{-1} q^{-1} x)}
       {\sY_2(q^{-2} x)}:
 +
 :\frac{\sY_3(\mu_3^{-1} x)\sY_3(\mu_3^{-1} q^{-1} x)}
       {\sY_2(q^{-2} x)}:
 +
 :\frac{\sY_4(\mu_4^{-1} x)\sY_4(\mu_4^{-1} q^{-1} x)}
       {\sY_2(q^{-2} x)}:
 \, , 
\end{align}
\begin{align}
 T_2^-(x) & = \,
 :\frac{\sY_1(\mu_1^{-1} q^{-1} x) \sY_3(\mu_3^{-1} q^{-1} x)
        \sY_4(\mu_4^{-1} q^{-1} x)}
       {\sY_2(q^{-2} x)^2}:
 +
 :\frac{\sY_2(q^{-1} x)}
       {\sY_1(\mu_1^{-1} q^{-1} x) \sY_1(\mu_1^{-1} q^{-2} x)}:
 \nonumber \\[.5em]
 &
 +
 :\frac{\sY_2(q^{-1} x)}
       {\sY_3(\mu_3^{-1} q^{-1} x) \sY_3(\mu_3^{-1} q^{-2} x)}:
 +
 :\frac{\sY_2(q^{-1} x)}
       {\sY_4(\mu_4^{-1} q^{-1} x) \sY_4(\mu_4^{-1} q^{-2} x)}:
 +
 :\frac{\sY_3(\mu_3^{-1} q^{-1} x) \sY_4(\mu_4^{-1} q^{-1} x)}
       {\sY_1(\mu_1^{-1} q^{-2} x) \sY_2(q^{-2} x)}:
 \nonumber \\[.5em]
 &
 +
 :\frac{\sY_1(\mu_1^{-1} q^{-1} x) \sY_4(\mu_4^{-1} q^{-1} x)}
       {\sY_3(\mu_3^{-1} q^{-2} x) \sY_2(q^{-2} x)}: 
 +
 :\frac{\sY_1(\mu_1^{-1} q^{-1} x) \sY_3(\mu_3^{-1} q^{-1} x)}
       {\sY_4(\mu_4^{-1} q^{-2} x) \sY_2(q^{-2} x)}: 
 +
 :\frac{\sY_1(\mu_1^{-1} q^{-1} x)}
       {\sY_3(\mu_3^{-1} q^{-2} x) \sY_4(\mu_4^{-1} q^{-2} x)}:
 \nonumber \\[.5em]
 &
 +
 :\frac{\sY_3(\mu_1^{-1} q^{-1} x)}
       {\sY_1(\mu_1^{-1} q^{-2} x) \sY_4(\mu_4^{-1} q^{-2} x)}:
 +
 :\frac{\sY_4(\mu_4^{-1} q^{-1} x)}
       {\sY_1(\mu_1^{-1} q^{-2} x) \sY_3(\mu_3^{-1} q^{-2} x)}:
 \nonumber \\[.5em]
 &
 +
 :\frac{\sY_2(q^{-2} x)}
       {\sY_1(\mu_1^{-1} q^{-2} x)
        \sY_3(\mu_3^{-1} q^{-2} x)
        \sY_4(\mu_4^{-1} q^{-2} x)}:
 + \, \sY_2(q^{-3}x)^{-1}
 \, .
\end{align}
In the classical limit, this $T_2$-operator is reduced to the character of 28 dimensional representation of $SO(8)$.
The $\msS$-factor appears at the zero weight terms.

\section{Affine quiver W-algebra}
\label{sec:affine_alg}

Applying our construction to affine quiver gauge theory, we can define a new W-algebra, associated to affine Lie algebra.
Let us consider the simplest affine quiver $\Gamma = \widehat{A}_0$, corresponding to 4d $\CalN = 2^*$ (5d $\CalN = 1^*$) theory, whose deformation of Cartan matrix is given by
\begin{align}
 c^{[n]} & =
 1 + q^{-n} - \mu^{-n} - \mu^{n} q^{-n}
 \ \stackrel{n \to 0}{\longrightarrow} \ 0
 \, .
\end{align}
This Cartan matrix can be nontrivial only in the presence of the multiplicative adjoint mass parameter $\mu \, (\neq 1)$.
This means that the mass deformation of the Cartan matrix is essential for the affine quiver, and this deformation goes beyond Frenkel--Reshetikhin's construction of W-algebra~\cite{Frenkel:1996,Frenkel:1997}.
%From the gauge theory point of view, the limits $\mu \to 1$ and $\mu \to \infty$ would be interesting to study, since they correspond to $\CalN=4$ and pure $\CalN=2$ theory.
%But how such a limit works is not obvious from this algebraic point of view.
The free field operators for $\sY$ and $\sS$ fields obey the commutation relation
\begin{align}
 \Big[ y_{1,n}, y_{1,m} \Big] & =
 - \frac{1}{n} \frac{(1-q_1^n)(1-q_2^n)}{(1-\mu^n)(1-\mu^{-n}q^n)}
 \delta_{n+m,0}
 \, ,
\end{align}
and
\begin{align}
 \Big[ s_{1,n}, s_{1,m} \Big] & =
 - \frac{1}{n} \frac{1 - q_1^n}{1 - q_2^{-n}}
 (1-\mu^{-n})(1-\mu^n q^{-n}) \, \delta_{n+m,0}
 \, .
\end{align}
The $s$-oscillator becomes trivial in the limit $\mu \to 1$ (and also $\mu \to q$), which means that the Nekrasov factor becomes trivial in $\CalN=4$ theory, while the $y$-oscillator becomes singular in this limit.

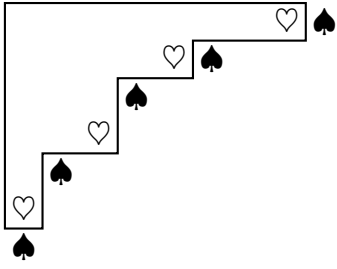
\begin{figure}[t]
 \begin{center}

  \begin{tikzpicture}[thick]

%   \node at (-1,-1.25) {$\lambda$ \ $=$};
   
   \draw (0,0) -- (4,0) -- (4,-.5) -- (2.5,-.5) -- (2.5,-1) -- (1.5,-1) -- (1.5,-2.) -- (1,-2.) -- (.5,-2.) -- (.5,-3.) -- (0,-3.) -- cycle ;

   \node at (4.25,-.25) {$\spadesuit$};
   \node at (2.75,-.75) {$\spadesuit$};
   \node at (1.75,-1.25) {$\spadesuit$};
   \node at (.75,-2.25) {$\spadesuit$};
   \node at (.25,-3.25) {$\spadesuit$};

   \node at (3.75,-.25) {$\heartsuit$};
   \node at (2.25,-.75) {$\heartsuit$};
   \node at (1.25,-1.75) {$\heartsuit$};
   \node at (.25,-2.75) {$\heartsuit$};
   
  \end{tikzpicture}
  
 \end{center}
 \caption{The outer and inner boundaries of the partition, $\partial_+ \lambda$ and $\partial_- \lambda$, denote the to-be-added and to-be-removed boxes, $\spadesuit$ and $\heartsuit$, respectively. The number of $\spadesuit$ is always bigger than $\heartsuit$ by one.}
 \label{fig:partition}
\end{figure}

The $T$-operator in this case is given by the iWeyl reflection as well, but it does not terminate within a finite sum: the affine character needs an infinite sum parametrized by the partition
\begin{align}
 T_1(x) & =
 \sY_1(x) \, + \fq \,
 \msS(\mu^{-1})
 : \sY_1(q^{-1}x)^{-1} \sY_1(\mu^{-1} x) \sY_1(\mu q^{-1} x) :
 + \cdots
 \nonumber \\[.5em]
 & =
 \sum_\lambda \fq^{|\lambda|}
 Z_\lambda^{\widehat{A}_0}(\tilde{q_1},\tilde{q_2},\tilde{\mu})
 :
 \prod_{s \in \partial_+ \lambda} \sY_1(qx/\tilde{x}(s))
 \prod_{s \in \partial_- \lambda} \sY_1(x/\tilde{x}(s))^{-1}
 :
\end{align}
where the coupling constant $\fq$ is now explicitly appearing because the Cartan matrix $c = (0)$ is not invertible for affine quiver $\Gamma = \widehat{A}_0$.
$\partial_+ \lambda$ and $\partial_- \lambda$ are the outer and inner boundary of the partition $\lambda$, as shown in Fig.~\ref{fig:partition}, and we define
\begin{align}
 \tilde{x}(s)
 & = (\mu^{-1} q)^{s_1-1} \mu^{s_2-1} q
 \, ,
\end{align}
which reads the dual variables
\begin{align}
 \tilde{q}_1 = \mu^{-1} q
 \, , \qquad
 \tilde{q}_2 = \mu
 \, , \qquad
 \tilde{\mu} = q_2
 \, , \qquad
 \tilde{\nu} = q
 \, .
\end{align}
We remark that $\tilde{q} := \tilde{q}_1 \tilde{q}_2 = q$.
The Nekrasov instanton function $Z_\lambda^{\widehat{A}_0}(\tilde{q_1},\tilde{q_2},\tilde{\mu})$ is again for $\widehat{A}_0$ quiver, but $U(1)$ gauge theory %evaluated with dual variables
\begin{align}
 Z_\lambda^{\widehat{A}_0}(q_1,q_2,\mu)
 & =
 \prod_{(s_1,s_2) \in \lambda}
 \frac{(1-\mu \, q_1^{l(s)+1}q_2^{-a(s)})
       (1-\mu^{-1}q_1^{l(s)}q_2^{-a(s)-1})}
      {(1-q_1^{l(s)+1}q_2^{-a(s)})(1-q_1^{l(s)}q_2^{-a(s)-1})}
\end{align}
where the arm and leg lengths are defined: $a(s) = \lambda_{s_1} - s_2$, $l(s) = \lambda^t_{s_2} - s_1$.
The higher degree current is given as a summation over multiple partition.
This dual expansion is naturally explained using the 8-dimensional setup~\cite{Nekrasov:2015wsu}.

\section{Quiver elliptic W-algebra}
\label{sec:elliptic_alg}

We have constructed the $q$-deformed W-algebra W$_{q_1,q_2}(\Gamma)$ associated with $\Gamma$-quiver gauge theory in 5d, which corresponds to trigonometric (relativistic) integrable system.
According to such a connection with integrable systems, it is natural to consider its elliptic deformation using 6d gauge theory compactified on the elliptic curve $\BR^4 \times T^2$~\cite{Kimura:2016dys}.
See also \cite{Tan:2013xba,Koroteev:2015dja,Iqbal:2015fvd,Nieri:2015dts,Mironov:2015thk,Koroteev:2016znb,Mironov:2016cyq,Mironov:2016yue,Tan:2016cky} for recent results along this direction.

In this case the spectral parameter is periodic
\begin{align}
 x \simeq p x
 \label{eq:periodicity_6d}
\end{align}
where $\tau$ is the modulus of $T^2$ that defines $p = e^{2\pi i \tau} \in \BC^\times$.
One needs to impose this periodicity on the index functor.
Recalling that the 5d gauge theory partition function on $\BR^4 \times S^1$ is given by the index along $S^1$ defined
\begin{align}
 \mathbb{I} \left[ \sum_\alpha x_\alpha \right] & =
 \prod_\alpha \left( 1 - x_\alpha^{-1} \right)
 \, ,
\end{align}
the elliptic index (equivariant elliptic genus on $T^2$) yields the 6d partition function for $\BR^4 \times T^2$
\begin{align}
 \mathbb{I}_p \left[ \sum_\alpha x_\alpha \right] & =
 \prod_\alpha \theta(x_\alpha^{-1};p)
 \label{eq:ind-el}
\end{align}
where the Jacobi theta function is given by%
\footnote{%
In this article we apply the Dolbeault convention, which obeys
\begin{align}
 \mathbb{I}\left[x^{-1}\right] = -x^{-1} \mathbb{I}[x]
 \, , \qquad
 \mathbb{I}_p\left[x^{-1}\right] = -x^{-1} \mathbb{I}_p[x]
 \, .
\end{align}
This convention can be converted into the Dirac index,
\begin{align}
 \mathbb{I} \left[ \sum_\alpha x_\alpha \right] =
 \prod_\alpha \left( x_\alpha^{1/2} - x_\alpha^{-1/2} \right) 
 \, , \qquad
 \mathbb{I}_p \left[ \sum_\alpha x_\alpha \right] =
 \prod_\alpha \theta_1(x_\alpha;p)
\end{align}
with
\begin{align}
 \mathbb{I}\left[x^{-1}\right] = - \mathbb{I}\left[x\right]
 \, , \qquad
 \mathbb{I}_p\left[x^{-1}\right] = - \mathbb{I}_p\left[x\right]
 \, .
\end{align}
Under the conformal condition~\eqref{eq:anom_free_cond}, these two conventions are consistent.
}
\begin{align}
 \theta(x;p)
 & = (x;p)_\infty (px^{-1};p)_\infty
 \nonumber \\
 & = \exp
 \left(
  - \sum_{n \neq 0} \frac{1}{n} \frac{x^n}{1-p^n}
 \right)
 \, .
 \label{eq:int-el-exp}
\end{align}
We assume $|p| < 1$ and this elliptic system is reduced to 5d theory in the limit $p \to 0$.

Since the elliptic index \eqref{eq:ind-el} involves both positive and negative powers due to the expansion \eqref{eq:int-el-exp}, we need to introduce ``positive'' and ``negative'' oscillators~\cite{Clavelli:1973uk,Saito:2014PRIMS,Kimura:2016dys}
\begin{align}
 \Big[ y_{i,n}^{(\pm)} , y_{j,m}^{(\pm)} \Big]
 & =
 \mp \frac{1}{n} \frac{(1-q_1^{\pm n})(1-q_2^{\pm n})}{1-p^{\pm n}} \,
 \tilde{c}_{ij}^{[\mp n]} \delta_{n+m,0}
 \, ,
 \label{eq:y-field-el}
 \\[1em]
 \left[ s_{i,n}^{(\pm)} , s_{j,m}^{(\pm)} \right]
 & =
 \mp \frac{1}{n} \frac{1-q_1^{\pm n}}{(1-p^{\pm n})(1-q_2^{\mp n})} \,
 c_{ij}^{[\pm n]} \delta_{n+m,0}
 \, ,
 \label{eq:s-field-el}
\end{align}
and thus
\begin{align}
 \sY_i(x) & = q_1^{\tilde{\rho}_i}
 :
 \exp \left( y_{i,0}
 + \sum_{n \neq 0} y_{i,n}^{(+)} x^{-n}
 + \sum_{n \neq 0} y_{i,n}^{(-)} x^{+n}
 \right)
 :
 \, , \\
 S_i(x) & = \,
 :
 \exp \left(
 s_{i,0} \log x + \tilde{s}_{i,0}
 + \sum_{n \neq 0} s_{i,n}^{(+)} x^{-n}
 + \sum_{n \neq 0} s_{i,n}^{(-)} x^{+n}
 \right)
 :
 \, .
\end{align}
Then the remaining construction is completely parallel with 5d theory:
The $T$-operator is generated by the iWeyl reflection with the $\sY$-operator, and the $Z$-state is constructed using the screening charge, built with the elliptic oscillators \eqref{eq:y-field-el} and \eqref{eq:s-field-el}.
In this procedure, the whole rational functions are replaced with the elliptic function.
For example, the scalar factor \eqref{eq:S-factor} due to the $\sY$ and $\sA$ OPE becomes
\begin{align}
 \msS(x) & =
 \frac{\theta(q_1 x;p) \theta(q_2 x;p)}
      {\theta(q x;p)\theta(x;p)}
 \, ,
\end{align}
and the commutation relation between the $\sY$ and $S$ operators is given by
\begin{align}
 \Big[ \sY_i(x), S_j(x') \Big] & =
 \delta_{i,j}
 \frac{\theta(q_1^{-1};p)}{(p;p)_\infty^2}
 \delta \left( q_1 \frac{x'}{x} \right)
 :\sY_i (x) S_j(x'):
\end{align}
which becomes zero in the limit $q_1 \to 1$.

\begin{figure}[t]
 \centering
 \begin{tikzpicture}[thick,scale=.8]

  \draw (0,0) -- (3,0)
  arc [start angle = -90, end angle = 90, x radius = .3, y radius = 1]
  -- (0,2)
  arc [start angle = 90, end angle = -270, x radius = .3, y radius = 1];

%  \draw [dashed] (3,0)
%  arc [start angle = -90, end angle = -270, x radius = .3, y radius = 1];

  \draw (5,0)
  arc [start angle = -90, end angle = 90, x radius = 1, y radius = 1]
  arc [start angle = 90, end angle = -270, x radius = .3, y radius = 1];

  \draw (-2,0)
  arc [start angle = -90, end angle = 90, x radius = .3, y radius = 1]
  arc [start angle = 90, end angle = 270, x radius = 1, y radius = 1];

  \draw [dotted] (-2,0)
  arc [start angle = -90, end angle = -270, x radius = .3, y radius = 1];

  \node at (-1,1) {$\times$};
  \node at (4,1) {$\times$};

  \node at (-1,-1.5) {$\times$};
  \node at (4,-1.5) {$\times$};  

%  \node (x1) at (.8,1.5) {$\bullet$};
%  \node (x2) at (2.6,.5) {$\bullet$};  

  \node at (-1.6,-1.5) [left] {$\bra{1}$};
  \node at (4.6,-1.5) [right] {$\ket{1}$};  
  \node at (1.5,-1.5)
  {$\displaystyle \prod_{\mathring{x} \in \CalX_0}^\succ
    \sS_{\si(\mathring{x}),\mathring{x}}^\text{5d} $};

  \node at (7,1) {$=$};

  \begin{scope}[shift={(11,0)}]

   \draw (0,0) -- (3,0)
   arc [start angle = -90, end angle = 90, x radius = 1, y radius = 1]
   -- (0,2)
   arc [start angle = 90, end angle = -270, x radius = .3, y radius = 1];
   
  \draw (-2,0)
  arc [start angle = -90, end angle = 90, x radius = .3, y radius = 1]
  arc [start angle = 90, end angle = 270, x radius = 1, y radius = 1];

  \draw [dotted] (-2,0)
  arc [start angle = -90, end angle = -270, x radius = .3, y radius = 1];

  \node at (-1,1) {$\times$};
  \node at (-1,-1.5) {$\times$};

   \node at (-1.6,-1.5) [left] {$\bra{1}$};
   \node at (1.7,-1.5)
   {$\displaystyle \ket{Z^\text{5d}}$};
   
  \end{scope}

  \begin{scope}[shift={(0,-6.5)}]

  \draw (-.5,0) -- (1.5,0)
  arc [start angle = -90, end angle = 90, x radius = .3, y radius = 1]
  -- (-.5,2)
  arc [start angle = 90, end angle = -270, x radius = .3, y radius = 1];

  \draw (2.25,0) -- ++(2,0)
  arc [start angle = -90, end angle = 90, x radius = .3, y radius = 1]
  -- (2.25,2)
  arc [start angle = 90, end angle = -270, x radius = .3, y radius = 1];

  \draw (5,0) -- ++(2,0)
  arc [start angle = -90, end angle = 90, x radius = .3, y radius = 1]
  -- (5,2)
  arc [start angle = 90, end angle = -270, x radius = .3, y radius = 1];
   
  \draw (8.5,0)
  arc [start angle = -90, end angle = 90, x radius = 1, y radius = 1]
  arc [start angle = 90, end angle = -270, x radius = .3, y radius = 1];

  \draw (-2,0)
  arc [start angle = -90, end angle = 90, x radius = .3, y radius = 1]
  arc [start angle = 90, end angle = 270, x radius = 1, y radius = 1];

  \draw [dotted] (-2,0)
  arc [start angle = -90, end angle = -270, x radius = .3, y radius = 1];

   \node at (-1.2,1) {$\cdots$};
   \node at (7.8,1) {$\cdots$};

   \node at (-1.2,2.5) {$\cdots$};
   \node at (7.8,2.5) {$\cdots$};   

   \node at (10.5,1) {$=$};
   \node at (10.5,-1.7) {$=$};   

   \node at (-1.2,-1.7) {$\times$};
   \node at (7.8,-1.7) {$\times$};

   \node at (-1.6,-1.7) [left] {$\bra{1}$};
   \node at (8.3,-1.7) [right] {$\ket{1}$};  
   \node at (3.25,-1.7)
   {$\displaystyle \prod_{\mathring{x} \in \CalX_0}^\succ
     \sS_{\si(\mathring{x}),\mathring{x}}^\text{6d} $};

   \draw
   [decorate,decoration={brace,amplitude=10pt,raise=-4pt}]
   (7.8,-.5) -- (-1.3,-.5) node [black,midway,yshift=-0.6pt] {};
   
%   \node at (3.25,-3) {\eqref{eq:Z-conf_block}};

   \node at (.5,2.5) {$p^{-1}x$};
   \node at (3.25,2.5) {$x$};
   \node at (6,2.5) {$p x$};   
   
  \end{scope}

  \begin{scope}[shift={(13,-6.5)}]
   
   \draw (1,0) -- ++(1.5,0)
   arc [start angle = -90, end angle = 90, x radius = .3, y radius = 1]
   -- (1,2)
   arc [start angle = 90, end angle = -270, x radius = .3, y radius = 1];

   \draw [thick] (1,1)
   circle [x radius = .3, y radius = 1];

   \draw [thick] (2.5,0)
   arc [start angle = -90, end angle = 90, x radius = .3, y radius = 1];

   \draw [thick,dotted] (2.5,0)
   arc [start angle = -90, end angle = -270, x radius = .3, y radius = 1];

   \node at (-1.5,1) {Tr};

   \node at (-.15,1) {$p^{L_0}$ $\times$};

   \draw (-.8,-.5) -- (-1,-.5) -- (-1,2.5) -- (-.8,2.5);
   \draw (3,-.5) -- (3.2,-.5) -- (3.2,2.5) -- (3,2.5);

   \node at (1.3,-1.7) {\eqref{eq:Tr_formula}};

  \end{scope}
  
 \end{tikzpicture}
 \caption{Conformal blocks as the partition function of five-dimensional (top) and six-dimensional theory (bottom). The six-dimensional block has two equivalent expressions.}
 \label{fig:conf_block}
\end{figure}

We can obtain the non-extended partition function from the $Z$-state as a correlator of the elliptic algebra as well as \eqref{eq:Z-correlator}.
In this case, the elliptic correlator can be also written in terms of 5d operators%
\footnote{%
To assure the modularity of (non-extended) gauge theory partition function, we need to assign the anomaly-free condition for 6d theory
\begin{align}
 \sum_{j \in \Gamma_0 }c_{ij} \, n_j = n_i^\text{flavor}
 \quad
 \text{for}
 \quad
 \forall i \in \Gamma_0
 \, .
 \label{eq:anom_free_cond}
\end{align}
This is equivalent to the conformal condition in 4d quiver gauge theory.
}
\begin{align}
 Z(t=0) =
 \langle 1 | \prod_{\mathring{x}\in\CalX_0}^{\succ} \sS^\text{6d}_{\si(\mathring{x})}(\mathring{x}) | 1 \rangle
 =
 \Tr
 \left[
 p^{L_0}
 \prod_{\mathring{x}\in\CalX_0}^{\succ}
 \sS^\text{5d}_{\si(\mathring{x})}(\mathring{x})
 \right]
 \label{eq:Tr_formula}
\end{align}
where the energy operator is defined
\begin{align}
 L_0 & =
 \sum_{i \in \Gamma_0} \sum_{m = 1}^\infty
 m \, t_{i,m} \frac{\partial}{t_{i,m}}
 \, .
\end{align}
The meaning of this trace formula is as follows.
One possibility to impose the periodicity \eqref{eq:periodicity_6d} is doubling the free fields.
Another possible way is to change the definition of the correlator to the trace, as shown in the bottom of Fig.~\ref{fig:conf_block}, as well as the usual CFT correlator on a torus.
They are actually two equivalent methods to compute the torus correlation function~\cite{Clavelli:1973uk,Kimura:2016dys}.
From this point of view, the $Z$-state itself is not necessarily modular invariant, while the torus correlator should be.

%\subsection{$A_r$ quiver}

%\rem{Examples: $A_1$-quiver~\cite{Nieri:2015dts} \& $A_2$-quiver}

Let us show the generating current of elliptic W-algebra with $A_r$ linear quiver examples.
The generating current itself is constructed by the $\sY$-operator in the same way as before.
The elliptic W($A_1$) generating current is given by
\begin{align}
 T_1(x)
 & =
 \sY_1(x) + \frac{1}{\sY_1(q^{-1} x)}
 \, ,
\end{align}
as discussed in Sec.~\ref{sec:A1}, and W($A_2$) currents are given by
\begin{align}
 T_1(x) & =
 \sY_1 (x) \, +
 :\frac{\sY_2(\mu^{-1}x)}{\sY_1(q^{-1}x)}:
 + \, \frac{1}{\sY_2(\mu^{-1} q^{-1} x)}
 \, , \\
 T_2(x) & =
 \sY_2 (x) \, +
 :\frac{\sY_1(\mu q^{-1} x)}{\sY_2(q^{-1}x)}:
 + \, \frac{1}{\sY_1(\mu q^{-2} x)}
 \, ,
\end{align}
where $\mu = \mu_{1 \to 2} = \mu_{2 \to 1}^{-1} q$, as shown in Sec.~\ref{sec:A2}.
The construction is also parallel for the linear quiver $A_r$ shown in Sec.~\ref{sec:Ar}.

These generating currents exhibit a similar OPE relation.
For $A_1$ quiver we have~\cite{Nieri:2015dts}
\begin{align}
 &
 f_{11}\left(\frac{y}{x}\right) T_1(x) T_1(y)
 - f_{11}\left(\frac{x}{y}\right) T_1(y) T_1(x)
 \nonumber \\
 & \hspace{8em}
 =
 - \frac{\theta(q_1;p)\theta(q_2;p)}{(p;p)_\infty^2 \theta(q;p)}
 \left(
 \delta \left(\frac{qy}{x}\right)
 - \delta \left(\frac{qx}{y}\right)
 \right)
 \label{eq:eVir_OPE}
\end{align}
and for $A_2$ quiver we have
\begin{align}
 &
 f_{11} \left(\frac{y}{x}\right) T_1(x) T_1(y)
 - f_{11} \left(\frac{x}{y}\right) T_1(y) T_1(x)
 \nonumber \\
 & \hspace{8em}
 =
 - \frac{\theta(q_1;p)\theta(q_2;p)}{(p;p)_\infty^2 \theta(q;p)}
 \left(
 \delta\left(q\frac{y}{x}\right) T_2(\mu^{-1} x)
 - \delta\left(q^{-1}\frac{y}{x}\right) T_2(\mu^{-1} q x)
 \right)
 \\[.5em]
 &
 f_{12}\left(\frac{y}{x}\right) T_1(x) T_2(y)
 - f_{21}\left(\frac{x}{y}\right) T_2(y) T_1(x)
  \nonumber \\
 & \hspace{8em}
 =
 - \frac{\theta(q_1;p)\theta(q_2;p)}{(p;p)_\infty^2 \theta(q;p)}
 \left(
 \delta\left(\mu q \frac{y}{x}\right)
 - \delta\left(\mu q^{-2}\frac{y}{x}\right)
 \right)
 \\[.5em]
 &\
 f_{22} \left(\frac{y}{x}\right) T_2(x) T_2(y)
 - f_{22} \left(\frac{x}{y}\right) T_2(y) T_2(x)
 \nonumber \\
 & \hspace{8em} 
 =
 - \frac{\theta(q_1;p)\theta(q_2;p)}{(p;p)_\infty^2 \theta(q;p)}
 \left(
 \delta\left(q\frac{y}{x}\right) T_1(\mu q^{-1} x)
 - \delta\left(q^{-1}\frac{y}{x}\right) T_1(\mu x)
 \right)
\end{align}
where
%\begin{align}
% \sY_i(x) \sY_j(y)
% & =
% f_{ij}\left(\frac{y}{x}\right)^{-1}
% : \sY_i(x) \sY_j(y) :
%\end{align}
%with
\begin{align}
 f_{ij} (x) & =
 \exp
 \left(
 \sum_{n \neq 0}^\infty \frac{(1-q_1^n)(1-q_2^n)}{n (1-p^n)} \,
 \tilde{c}_{ij}^{[-n]} x^n
 \right)
 \, .
\end{align}
These OPE relations characterize the algebraic relation for the elliptic W-algebra generators.

\subsection*{Acknowledgements}

The author would like to thank V.~Pestun for fruitful collaboration which materializes this article and useful comments on the draft.
This article is based on a talk presented at \href{http://www.ams.org/meetings/amsconf/symposia/symposia-2016}{2016 AMS von Neumann Symposium}: Topological Recursion and its Influence in Analysis, Geometry, and Topology, July 4--8, 2016, Hilton Charlotte University Place, Charlotte, NC.
The author is grateful to the symposium organizers for giving the opportunity to present our work in a stimulating atmosphere.
This work was supported in part by Keio Gijuku Academic Development Funds, JSPS Grant-in-Aid for Scientific Research (No.~JP17K18090), the MEXT-Supported Program for the Strategic Research Foundation at Private Universities ``Topological Science'' (No.~S1511006), JSPS Grant-in-Aid for Scientific Research on Innovative Areas ``Topological Materials Science'' (No.~JP15H05855), and ``Discrete Geometric Analysis for Materials Design'' (No.~JP17H06462).

%%%%%%%%%% References %%%%%%%%%%

\bibliographystyle{utphysurl} \bibliography{wquiver}

\providecommand{\href}[2]{#2}\begingroup\raggedright\begin{thebibliography}{10}

\bibitem{Seiberg:1994rs}
N.~Seiberg and E.~Witten, ``{Electric-magnetic duality, monopole condensation,
  and confinement in $N=2$ supersymmetric Yang--Mills theory},''
  \href{http://dx.doi.org/10.1016/0550-3213(94)90124-4}{{\em Nucl. Phys.}
  {\bfseries B426} (1994) 19--52},
  \href{http://arxiv.org/abs/hep-th/9407087}{{\ttfamily arXiv:hep-th/9407087
  [hep-th]}}.
[Erratum: Nucl. Phys. \textbf{B430} (1994) 485].
%%CITATION = HEP-TH/9407087;%%.

\bibitem{Seiberg:1994aj}
N.~Seiberg and E.~Witten, ``{Monopoles, duality and chiral symmetry breaking in
  $N=2$ supersymmetric QCD},''
  \href{http://dx.doi.org/10.1016/0550-3213(94)90214-3}{{\em Nucl. Phys.}
  {\bfseries B431} (1994) 484--550},
\href{http://arxiv.org/abs/hep-th/9408099}{{\ttfamily arXiv:hep-th/9408099
  [hep-th]}}.
%%CITATION = HEP-TH/9408099;%%.

\bibitem{Gorsky:1995zq}
A.~Gorsky, I.~Krichever, A.~Marshakov, A.~Mironov, and A.~Morozov,
  ``{Integrability and Seiberg--Witten exact solution},''
  \href{http://dx.doi.org/10.1016/0370-2693(95)00723-X}{{\em Phys. Lett.}
  {\bfseries B355} (1995) 466--474},
\href{http://arxiv.org/abs/hep-th/9505035}{{\ttfamily arXiv:hep-th/9505035
  [hep-th]}}.
%%CITATION = HEP-TH/9505035;%%.

\bibitem{Martinec:1995by}
E.~J. Martinec and N.~P. Warner, ``{Integrable systems and supersymmetric gauge
  theory},'' \href{http://dx.doi.org/10.1016/0550-3213(95)00588-9}{{\em Nucl.
  Phys.} {\bfseries B459} (1996) 97--112},
\href{http://arxiv.org/abs/hep-th/9509161}{{\ttfamily arXiv:hep-th/9509161
  [hep-th]}}.
%%CITATION = HEP-TH/9509161;%%.

\bibitem{Donagi:1995cf}
R.~Donagi and E.~Witten, ``{Supersymmetric Yang--Mills theory and integrable
  systems},'' \href{http://dx.doi.org/10.1016/0550-3213(95)00609-5}{{\em Nucl.
  Phys.} {\bfseries B460} (1996) 299--334},
\href{http://arxiv.org/abs/hep-th/9510101}{{\ttfamily arXiv:hep-th/9510101
  [hep-th]}}.
%%CITATION = HEP-TH/9510101;%%.

\bibitem{Seiberg:1996nz}
N.~Seiberg and E.~Witten, ``{Gauge dynamics and compactification to
  three-dimensions},'' in {\em {The mathematical beauty of physics: A memorial
  volume for Claude Itzykson}}, vol.~24 of {\em Adv. Ser. Math. Phys.},
  pp.~333--366.
\newblock World Scientific, 1997.
\newblock
\href{http://arxiv.org/abs/hep-th/9607163}{{\ttfamily arXiv:hep-th/9607163
  [hep-th]}}.
\newblock
%%CITATION = HEP-TH/9607163;%%.

\bibitem{Gorsky:1996hs}
A.~Gorsky, A.~Marshakov, A.~Mironov, and A.~Morozov, ``{$\mathcal{N}=2$
  supersymmetric QCD and integrable spin chains: Rational case $N_F < 2
  N_c$},'' \href{http://dx.doi.org/10.1016/0370-2693(96)00480-7}{{\em Phys.
  Lett.} {\bfseries B380} (1996) 75--80},
\href{http://arxiv.org/abs/hep-th/9603140}{{\ttfamily arXiv:hep-th/9603140
  [hep-th]}}.
%%CITATION = HEP-TH/9603140;%%.

\bibitem{Nekrasov:1996cz}
N.~Nekrasov, ``{Five dimensional gauge theories and relativistic integrable
  systems},'' \href{http://dx.doi.org/10.1016/S0550-3213(98)00436-2}{{\em Nucl.
  Phys.} {\bfseries B531} (1998) 323--344},
\href{http://arxiv.org/abs/hep-th/9609219}{{\ttfamily arXiv:hep-th/9609219
  [hep-th]}}.
%%CITATION = HEP-TH/9609219;%%.

\bibitem{Nekrasov:2012xe}
N.~Nekrasov and V.~Pestun, ``{Seiberg--Witten geometry of four dimensional
  $N=2$ quiver gauge theories},''
  \href{http://arxiv.org/abs/1211.2240}{{\ttfamily arXiv:1211.2240 [hep-th]}}.

\bibitem{Moore:1997dj}
G.~W. Moore, N.~Nekrasov, and S.~Shatashvili, ``{Integrating over Higgs
  branches},'' \href{http://dx.doi.org/10.1007/PL00005525}{{\em Commun. Math.
  Phys.} {\bfseries 209} (2000) 97--121},
  \href{http://arxiv.org/abs/hep-th/9712241}{{\ttfamily hep-th/9712241}}.

\bibitem{Nekrasov:2002qd}
N.~A. Nekrasov, ``{Seiberg--Witten prepotential from instanton counting},''
  \href{http://dx.doi.org/10.4310/ATMP.2003.v7.n5.a4}{{\em Adv. Theor. Math.
  Phys.} {\bfseries 7} (2004) 831--864},
  \href{http://arxiv.org/abs/hep-th/0206161}{{\ttfamily hep-th/0206161}}.

\bibitem{Nekrasov:2009zz}
N.~Nekrasov and S.~Shatashvili, ``{Bethe Ansatz and supersymmetric vacua},''
\href{http://dx.doi.org/10.1063/1.3149487}{{\em AIP Conf. Proc.} {\bfseries
  1134} (2009) 154--169}.
%%CITATION = APCPC,1134,154;%%.

\bibitem{Nekrasov:2009rc}
N.~A. Nekrasov and S.~L. Shatashvili,
  \href{http://dx.doi.org/10.1142/9789814304634_0015}{``{Quantization of
  Integrable Systems and Four Dimensional Gauge Theories},''} in {\em {XVIth
  International Congress on Mathematical Physics}}, pp.~265--289.
\newblock 2009.
\newblock
\href{http://arxiv.org/abs/0908.4052}{{\ttfamily arXiv:0908.4052 [hep-th]}}.
\newblock
%%CITATION = ARXIV:0908.4052;%%.

\bibitem{Nekrasov:2013xda}
N.~Nekrasov, V.~Pestun, and S.~Shatashvili, ``{Quantum geometry and quiver
  gauge theories},'' \href{http://arxiv.org/abs/1312.6689}{{\ttfamily
  arXiv:1312.6689 [hep-th]}}.

\bibitem{Frenkel:1998}
E.~Frenkel and N.~Reshetikhin,
  \href{http://dx.doi.org/10.1090/conm/248/03823}{``{The {$q$}-characters of
  representations of quantum affine algebras and deformations of
  {$\mathcal{W}$}-algebras},''} in {\em {Recent Developments in Quantum Affine
  Algebras and Related Topics}}, vol.~248 of {\em Contemp. Math.},
  pp.~163--205.
\newblock Amer. Math. Soc., 1999.
\newblock \href{http://arxiv.org/abs/math/9810055}{{\ttfamily math/9810055
  [math.QA]}}.

\bibitem{Hollowood:2003cv}
T.~J. Hollowood, A.~Iqbal, and C.~Vafa, ``{Matrix models, geometric engineering
  and elliptic genera},''
  \href{http://dx.doi.org/10.1088/1126-6708/2008/03/069}{{\em JHEP} {\bfseries
  0803} (2008) 069},
\href{http://arxiv.org/abs/hep-th/0310272}{{\ttfamily arXiv:hep-th/0310272
  [hep-th]}}.
%%CITATION = HEP-TH/0310272;%%.

\bibitem{Nekrasov:2015wsu}
N.~Nekrasov, ``{BPS/CFT correspondence: non-perturbative Dyson--Schwinger
  equations and $qq$-characters},''
  \href{http://dx.doi.org/10.1007/JHEP03(2016)181}{{\em JHEP} {\bfseries 1603}
  (2016) 181},
\href{http://arxiv.org/abs/1512.05388}{{\ttfamily arXiv:1512.05388 [hep-th]}}.
%%CITATION = ARXIV:1512.05388;%%.

\bibitem{Nekrasov:2016qym}
N.~Nekrasov, ``{BPS/CFT correspondence II: Instantons at crossroads, Moduli and
  Compactness Theorem},''
  \href{http://dx.doi.org/10.4310/ATMP.2017.v21.n2.a4}{{\em Adv. Theor. Math.
  Phys.} {\bfseries 21} (2017) 503--583},
\href{http://arxiv.org/abs/1608.07272}{{\ttfamily arXiv:1608.07272 [hep-th]}}.
%%CITATION = ARXIV:1608.07272;%%.

\bibitem{Bourgine:2015szm}
J.-E. Bourgine, Y.~Mastuo, and H.~Zhang, ``{Holomorphic field realization of
  SH$^c$ and quantum geometry of quiver gauge theories},''
  \href{http://dx.doi.org/10.1007/JHEP04(2016)167}{{\em JHEP} {\bfseries 1604}
  (2016) 167},
\href{http://arxiv.org/abs/1512.02492}{{\ttfamily arXiv:1512.02492 [hep-th]}}.
%%CITATION = ARXIV:1512.02492;%%.

\bibitem{Mironov:2015thk}
A.~Mironov, A.~Morozov, and Y.~Zenkevich, ``{On elementary proof of AGT
  relations from six dimensions},''
  \href{http://dx.doi.org/10.1016/j.physletb.2016.03.006}{{\em Phys. Lett.}
  {\bfseries B756} (2016) 208--211},
\href{http://arxiv.org/abs/1512.06701}{{\ttfamily arXiv:1512.06701 [hep-th]}}.
%%CITATION = ARXIV:1512.06701;%%.

\bibitem{Kim:2016qqs}
H.-C. Kim, ``{Line defects and 5d instanton partition functions},''
  \href{http://dx.doi.org/10.1007/JHEP03(2016)199}{{\em JHEP} {\bfseries 1603}
  (2016) 199},
\href{http://arxiv.org/abs/1601.06841}{{\ttfamily arXiv:1601.06841 [hep-th]}}.
%%CITATION = ARXIV:1601.06841;%%.

\bibitem{Bourgine:2016vsq}
J.-E. Bourgine, M.~Fukuda, Y.~Matsuo, H.~Zhang, and R.-D. Zhu, ``{Coherent
  states in quantum $\mathcal{W}_{1+\infty}$ algebra and qq-character for 5d
  Super Yang-Mills},'' \href{http://dx.doi.org/10.1093/ptep/ptw165}{{\em PTEP}
  {\bfseries 2016} (2016) 123B05},
\href{http://arxiv.org/abs/1606.08020}{{\ttfamily arXiv:1606.08020 [hep-th]}}.
%%CITATION = ARXIV:1606.08020;%%.

\bibitem{Mironov:2016cyq}
A.~Mironov, A.~Morozov, and Y.~Zenkevich, ``{Spectral duality in elliptic
  systems, six-dimensional gauge theories and topological strings},''
  \href{http://dx.doi.org/10.1007/JHEP05(2016)121}{{\em JHEP} {\bfseries 1605}
  (2016) 121},
\href{http://arxiv.org/abs/1603.00304}{{\ttfamily arXiv:1603.00304 [hep-th]}}.
%%CITATION = ARXIV:1603.00304;%%.

\bibitem{Mironov:2016yue}
A.~Mironov, A.~Morozov, and Y.~Zenkevich, ``{Ding--Iohara--Miki symmetry of
  network matrix models},''
  \href{http://dx.doi.org/10.1016/j.physletb.2016.09.033}{{\em Phys. Lett.}
  {\bfseries B762} (2016) 196--208},
\href{http://arxiv.org/abs/1603.05467}{{\ttfamily arXiv:1603.05467 [hep-th]}}.
%%CITATION = ARXIV:1603.05467;%%.

\bibitem{Awata:2016riz}
H.~Awata, H.~Kanno, T.~Matsumoto, A.~Mironov, A.~Morozov, A.~Morozov,
  Y.~Ohkubo, and Y.~Zenkevich, ``{Explicit examples of DIM constraints for
  network matrix models},''
  \href{http://dx.doi.org/10.1007/JHEP07(2016)103}{{\em JHEP} {\bfseries 1607}
  (2016) 103},
\href{http://arxiv.org/abs/1604.08366}{{\ttfamily arXiv:1604.08366 [hep-th]}}.
%%CITATION = ARXIV:1604.08366;%%.

\bibitem{Awata:2016mxc}
H.~Awata, H.~Kanno, A.~Mironov, A.~Morozov, A.~Morozov, Y.~Ohkubo, and
  Y.~Zenkevich, ``{Toric Calabi-Yau threefolds as quantum integrable systems.
  $\mathcal{R}$-matrix and $\mathcal{RTT}$ relations},''
  \href{http://dx.doi.org/10.1007/JHEP10(2016)047}{{\em JHEP} {\bfseries 1610}
  (2016) 047},
\href{http://arxiv.org/abs/1608.05351}{{\ttfamily arXiv:1608.05351 [hep-th]}}.
%%CITATION = ARXIV:1608.05351;%%.

\bibitem{Kimura:2017auj}
T.~Kimura, H.~Mori, and Y.~Sugimoto, ``{Refined geometric transition and
  $qq$-characters},'' \href{http://dx.doi.org/10.1007/JHEP01(2018)025}{{\em
  JHEP} {\bfseries 1801} (2018) 025},
\href{http://arxiv.org/abs/1705.03467}{{\ttfamily arXiv:1705.03467 [hep-th]}}.
%%CITATION = ARXIV:1705.03467;%%.

\bibitem{Kimura:2015rgi}
T.~Kimura and V.~Pestun, ``{Quiver W-algebras},''
\href{http://arxiv.org/abs/1512.08533}{{\ttfamily arXiv:1512.08533 [hep-th]}}.
%%CITATION = ARXIV:1512.08533;%%.

\bibitem{Kimura:2016dys}
T.~Kimura and V.~Pestun, ``{Quiver elliptic W-algebras},''
\href{http://arxiv.org/abs/1608.04651}{{\ttfamily arXiv:1608.04651 [hep-th]}}.
%%CITATION = ARXIV:1608.04651;%%.

\bibitem{Kimura:2017hez}
T.~Kimura and V.~Pestun, ``{Fractional quiver W-algebras},''
\href{http://arxiv.org/abs/1705.04410}{{\ttfamily arXiv:1705.04410 [hep-th]}}.
%%CITATION = ARXIV:1705.04410;%%.

\bibitem{Marshakov:2006ii}
A.~Marshakov and N.~Nekrasov, ``{Extended Seiberg--Witten Theory and Integrable
  Hierarchy},'' \href{http://dx.doi.org/10.1088/1126-6708/2007/01/104}{{\em
  JHEP} {\bfseries 0701} (2007) 104},
  \href{http://arxiv.org/abs/hep-th/0612019}{{\ttfamily hep-th/0612019}}.

\bibitem{Alday:2009aq}
L.~F. Alday, D.~Gaiotto, and Y.~Tachikawa, ``{Liouville Correlation Functions
  from Four-dimensional Gauge Theories},''
  \href{http://dx.doi.org/10.1007/s11005-010-0369-5}{{\em Lett. Math. Phys.}
  {\bfseries 91} (2010) 167--197},
  \href{http://arxiv.org/abs/0906.3219}{{\ttfamily arXiv:0906.3219 [hep-th]}}.

\bibitem{Wyllard:2009hg}
N.~Wyllard, ``{$A_{N-1}$ conformal Toda field theory correlation functions from
  conformal $\mathcal{N} = 2$ $SU(N)$ quiver gauge theories},''
  \href{http://dx.doi.org/10.1088/1126-6708/2009/11/002}{{\em JHEP} {\bfseries
  0911} (2009) 002}, \href{http://arxiv.org/abs/0907.2189}{{\ttfamily
  arXiv:0907.2189 [hep-th]}}.

\bibitem{Awata:2009ur}
H.~Awata and Y.~Yamada, ``{Five-dimensional AGT Conjecture and the Deformed
  Virasoro Algebra},'' \href{http://dx.doi.org/10.1007/JHEP01(2010)125}{{\em
  JHEP} {\bfseries 1001} (2010) 125},
\href{http://arxiv.org/abs/0910.4431}{{\ttfamily arXiv:0910.4431 [hep-th]}}.
%%CITATION = ARXIV:0910.4431;%%.

\bibitem{Teschner:2010je}
J.~Teschner, ``{Quantization of the Hitchin moduli spaces, Liouville theory,
  and the geometric Langlands correspondence I},''
  \href{http://dx.doi.org/10.4310/ATMP.2011.v15.n2.a6}{{\em Adv. Theor. Math.
  Phys.} {\bfseries 15} (2011) 471--564},
\href{http://arxiv.org/abs/1005.2846}{{\ttfamily arXiv:1005.2846 [hep-th]}}.
%%CITATION = ARXIV:1005.2846;%%.

\bibitem{Manabe:2015kbj}
M.~Manabe and P.~Su\l{l}kowski, ``{Quantum curves and conformal field
  theory},'' \href{http://dx.doi.org/10.1103/PhysRevD.95.126003}{{\em Phys.
  Rev.} {\bfseries D95} (2017) 126003},
\href{http://arxiv.org/abs/1512.05785}{{\ttfamily arXiv:1512.05785 [hep-th]}}.
%%CITATION = ARXIV:1512.05785;%%.

\bibitem{Katz:1997eq}
S.~Katz, P.~Mayr, and C.~Vafa, ``{Mirror symmetry and exact solution of 4-D
  $N=2$ gauge theories: 1.},''
  \href{http://dx.doi.org/10.4310/ATMP.1997.v1.n1.a2}{{\em Adv. Theor. Math.
  Phys.} {\bfseries 1} (1998) 53--114},
  \href{http://arxiv.org/abs/hep-th/9706110}{{\ttfamily hep-th/9706110}}.

\bibitem{Aharony:1997bh}
O.~Aharony, A.~Hanany, and B.~Kol, ``{Webs of $(p,q)$ five-branes,
  five-dimensional field theories and grid diagrams},''
  \href{http://dx.doi.org/10.1088/1126-6708/1998/01/002}{{\em JHEP} {\bfseries
  01} (1998) 002},
\href{http://arxiv.org/abs/hep-th/9710116}{{\ttfamily arXiv:hep-th/9710116
  [hep-th]}}.
%%CITATION = HEP-TH/9710116;%%.

\bibitem{Bao:2011rc}
L.~Bao, E.~Pomoni, M.~Taki, and F.~Yagi, ``{M5-Branes, Toric Diagrams and Gauge
  Theory Duality},'' \href{http://dx.doi.org/10.1007/JHEP04(2012)105}{{\em
  JHEP} {\bfseries 1204} (2012) 105},
\href{http://arxiv.org/abs/1112.5228}{{\ttfamily arXiv:1112.5228 [hep-th]}}.
%%CITATION = ARXIV:1112.5228;%%.

\bibitem{Aganagic:2013tta}
M.~Aganagic, N.~Haouzi, C.~Koz\c{c}az, and S.~Shakirov, ``{Gauge/Liouville
  Triality},''
\href{http://arxiv.org/abs/1309.1687}{{\ttfamily arXiv:1309.1687 [hep-th]}}.
%%CITATION = ARXIV:1309.1687;%%.

\bibitem{Aganagic:2014oia}
M.~Aganagic, N.~Haouzi, and S.~Shakirov, ``{$A_n$-Triality},''
\href{http://arxiv.org/abs/1403.3657}{{\ttfamily arXiv:1403.3657 [hep-th]}}.
%%CITATION = ARXIV:1403.3657;%%.

\bibitem{Aganagic:2015cta}
M.~Aganagic and N.~Haouzi, ``{ADE Little String Theory on a Riemann Surface
  (and Triality)},''
\href{http://arxiv.org/abs/1506.04183}{{\ttfamily arXiv:1506.04183 [hep-th]}}.
%%CITATION = ARXIV:1506.04183;%%.

\bibitem{Frenkel:1996}
E.~Frenkel and N.~Reshetikhin, ``{Quantum affine algebras and deformations of
  the {V}irasoro and {$\mathscr{W}$}-algebras},''
  \href{http://dx.doi.org/10.1007/BF02104917}{{\em Comm. Math. Phys.}
  {\bfseries 178} (1996) 237--264},
  \href{http://arxiv.org/abs/q-alg/9505025}{{\ttfamily q-alg/9505025}}.

\bibitem{Frenkel:1997}
E.~Frenkel and N.~Reshetikhin, ``{Deformations of {$\mathcal{W}$}-algebras
  associated to simple {L}ie algebras},'' {\em Comm. Math. Phys.} {\bfseries
  197} (1998) 1--32, \href{http://arxiv.org/abs/q-alg/9708006}{{\ttfamily
  q-alg/9708006 [math.QA]}}.

\bibitem{Nekrasov:2003rj}
N.~A. Nekrasov and A.~Okounkov,
  \href{http://dx.doi.org/10.1007/0-8176-4467-9_15}{``{Seiberg--Witten Theory
  and Random Partitions},''} in {\em The Unity of Mathematics}, P.~Etingof,
  V.~Retakh, and I.~M. Singer, eds., vol.~244 of {\em Progress in Mathematics},
  pp.~525--596.
\newblock Birkh\"auser Boston, 2006.
\newblock
\href{http://arxiv.org/abs/hep-th/0306238}{{\ttfamily arXiv:hep-th/0306238
  [hep-th]}}.
\newblock
%%CITATION = HEP-TH/0306238;%%.

\bibitem{Shiraishi:1995rp}
J.~Shiraishi, H.~Kubo, H.~Awata, and S.~Odake, ``{A Quantum deformation of the
  Virasoro algebra and the Macdonald symmetric functions},''
  \href{http://dx.doi.org/10.1007/BF00398297}{{\em Lett. Math. Phys.}
  {\bfseries 38} (1996) 33--51},
  \href{http://arxiv.org/abs/q-alg/9507034}{{\ttfamily q-alg/9507034}}.

\bibitem{Dotsenko:1984nm}
V.~S. Dotsenko and V.~A. Fateev, ``{Conformal algebra and multipoint
  correlation functions in 2D statistical models},''
\href{http://dx.doi.org/10.1016/0550-3213(84)90269-4}{{\em Nucl. Phys.}
  {\bfseries B240} (1984) 312--348}.
%%CITATION = NUPHA,B240,312;%%.

\bibitem{Dotsenko:1984ad}
V.~S. Dotsenko and V.~A. Fateev, ``{Four-point correlation functions and the
  operator algebra in 2D conformal invariant theories with central charge $c
  \le 1$},''
\href{http://dx.doi.org/10.1016/S0550-3213(85)80004-3}{{\em Nucl. Phys.}
  {\bfseries B251} (1985) 691--734}.
%%CITATION = NUPHA,B251,691;%%.

\bibitem{Kharchev:1992iv}
S.~Kharchev, A.~Marshakov, A.~Mironov, A.~Morozov, and S.~Pakuliak,
  ``{Conformal matrix models as an alternative to conventional multimatrix
  models},'' \href{http://dx.doi.org/10.1016/0550-3213(93)90595-G}{{\em Nucl.
  Phys.} {\bfseries B404} (1993) 717--750},
\href{http://arxiv.org/abs/hep-th/9208044}{{\ttfamily arXiv:hep-th/9208044
  [hep-th]}}.
%%CITATION = HEP-TH/9208044;%%.

\bibitem{Kostov:1992ie}
I.~K. Kostov, ``{Gauge invariant matrix model for the A-D-E closed strings},''
  \href{http://dx.doi.org/10.1016/0370-2693(92)91072-H}{{\em Phys. Lett.}
  {\bfseries B297} (1992) 74--81},
  \href{http://arxiv.org/abs/hep-th/9208053}{{\ttfamily hep-th/9208053}}.

\bibitem{Awata:1995zk}
H.~Awata, H.~Kubo, S.~Odake, and J.~Shiraishi, ``{Quantum $\mathcal{W}_N$
  algebras and Macdonald polynomials},''
  \href{http://dx.doi.org/10.1007/BF02102595}{{\em Commun. Math. Phys.}
  {\bfseries 179} (1996) 401--416},
\href{http://arxiv.org/abs/q-alg/9508011}{{\ttfamily q-alg/9508011 [math.QA]}}.
%%CITATION = Q-ALG/9508011;%%.

\bibitem{Tan:2013xba}
M.-C. Tan, ``{An M-Theoretic Derivation of a 5d and 6d AGT Correspondence, and
  Relativistic and Elliptized Integrable Systems},''
  \href{http://dx.doi.org/10.1007/JHEP12(2013)031}{{\em JHEP} {\bfseries 1312}
  (2013) 031},
\href{http://arxiv.org/abs/1309.4775}{{\ttfamily arXiv:1309.4775 [hep-th]}}.
%%CITATION = ARXIV:1309.4775;%%.

\bibitem{Koroteev:2015dja}
P.~Koroteev and A.~Sciarappa, ``{Quantum Hydrodynamics from Large-$n$
  Supersymmetric Gauge Theories},''
  \href{http://dx.doi.org/10.1007/s11005-017-0996-1}{{\em Lett. Math. Phys.}
  {\bfseries 108} (2018) 45--95},
\href{http://arxiv.org/abs/1510.00972}{{\ttfamily arXiv:1510.00972 [hep-th]}}.
%%CITATION = ARXIV:1510.00972;%%.

\bibitem{Iqbal:2015fvd}
A.~Iqbal, C.~Koz\c{c}az, and S.-T. Yau, ``{Elliptic Virasoro Conformal
  Blocks},'' \href{http://arxiv.org/abs/1511.00458}{{\ttfamily arXiv:1511.00458
  [hep-th]}}.

\bibitem{Nieri:2015dts}
F.~Nieri, ``{An elliptic Virasoro symmetry in 6d},''
  \href{http://dx.doi.org/10.1007/s11005-017-0986-3}{{\em Lett. Math. Phys.}
  {\bfseries 107} (2017) 2147--2187},
  \href{http://arxiv.org/abs/1511.00574}{{\ttfamily arXiv:1511.00574
  [hep-th]}}.

\bibitem{Koroteev:2016znb}
P.~Koroteev and A.~Sciarappa, ``{On Elliptic Algebras and Large-$n$
  Supersymmetric Gauge Theories},''
  \href{http://dx.doi.org/10.1063/1.4966641}{{\em J. Math. Phys.} {\bfseries
  57} (2016) 112302},
\href{http://arxiv.org/abs/1601.08238}{{\ttfamily arXiv:1601.08238 [hep-th]}}.
%%CITATION = ARXIV:1601.08238;%%.

\bibitem{Tan:2016cky}
M.-C. Tan, ``{Higher AGT Correspondences, W-algebras, and Higher Quantum
  Geometric Langlands Duality from M-Theory},''
\href{http://arxiv.org/abs/1607.08330}{{\ttfamily arXiv:1607.08330 [hep-th]}}.
%%CITATION = ARXIV:1607.08330;%%.

\bibitem{Clavelli:1973uk}
L.~Clavelli and J.~A. Shapiro, ``{Pomeron factorization in general dual
  models},''
\href{http://dx.doi.org/10.1016/0550-3213(73)90113-2}{{\em Nucl. Phys.}
  {\bfseries B57} (1973) 490--535}.
%%CITATION = NUPHA,B57,490;%%.

\bibitem{Saito:2014PRIMS}
Y.~Saito, ``{Elliptic Ding--Iohara Algebra and the Free Field Realization of
  the Elliptic Macdonald Operator},''
  \href{http://dx.doi.org/10.4171/PRIMS/139}{{\em Pub. Res. Inst. Math. Sci.}
  {\bfseries 50} (2014) 411--455},
  \href{http://arxiv.org/abs/1301.4912}{{\ttfamily arXiv:1301.4912 [math.QA]}}.

\end{thebibliography}\endgroup
%\bibliographystyle{../wquiver/utphysurl}
%\bibliography{../wquiver/wquiver}

\end{document}